\documentclass[secnumarabic, graphics,floatfix, nofootinbib,tightenlines,nobibnotes, aps, prl, 12pt]{revtex4}
\usepackage{graphicx}
\usepackage[english]{babel}
\usepackage[utf8]{inputenc}
\usepackage{amsmath,amssymb}
\usepackage{amsfonts}
\usepackage{multirow}

\def\be{\begin{equation}}
\def\ee{\end{equation}}
\def\bea{\begin{eqnarray}}
\def\eea{\end{eqnarray}}

\begin{document}

\title{A Maxwell-vector p-wave holographic superconductor in a particular background AdS black hole metric}

\author{Dan Wen$^{1,2}$}
\author{Hongwei Yu$^{1}$\footnote{hwyu@hunnu.edu.cn}}
\author{Qiyuan Pan$^{1}$}
\author{Kai Lin$^{2}$}
\author{Wei-Liang Qian$^{3,4}$}

\affiliation{1) Key Laboratory of Low Dimensional Quantum Structures and Quantum Control of Ministry of Education, Synergetic Innovation Center for Quantum Effects and Applications, and Department of Physics, Hunan Normal University, Changsha, Hunan, China}
\affiliation{2) Universidade Federal de Itajub\'a, Instituto de F\'isica e Qu\'imica, Itajub\'a, MG, Brasil} 
\affiliation{3) Escola de Engenharia de Lorena, Universidade de S\~ao Paulo, Lorena, SP, Brasil} 
\affiliation{4) Faculadade de Engenharia de Guaratinguet\'a, Universidade Estadual Paulista, Guaratinguet\'a, SP, Brasil}

\begin{abstract}
We study the p-wave holographic superconductor for AdS black holes with planar event horizon topology for a particular Lovelock gravity, in which the action is characterized by a self-interacting scalar field nonminimally coupled to the gravity theory which is labeled by an integer $k$.
As the Lovelock theory of gravity is the most general metric theory of gravity based on the fundamental assumptions of general relativity, it is a desirable theory to describe the higher dimensional spacetime geometry.
The present work is devoted to studying the properties of the p-wave holographic superconductor by including a Maxwell field which nonminimally couples to a complex vector field in a higher dimensional background metric.
In the probe limit, we find that the critical temperature decreases with the increase of the index $k$ of the background black hole metric, which shows that a larger $k$ makes it harder for the condensation to form.
We also observe that the index $k$ affects the conductivity and the gap frequency of the holographic superconductors.
\end{abstract}

\maketitle
\newpage

\section{Introduction}

The mechanism behind high-temperature superconductor is one of the unsolved mysteries, which inspires long-standing interest in theoretical physics. As the AdS/CFT correspondence provides a viable approach for strong coupling systems, the holographic superconductivity \cite{G11,C09,NHM14,D04} has quickly become one of the most influential topics in particle physics. 
The conjectured gauge/gravity duality allows one to associate a (d+1)-dimensional classical gravity system with a d-dimensional strongly coupled conformal field theory on its boundary \cite{J98,SIA98}. 
The superconductor and its phase transition, therefore, can be analyzed through the ``hair''/``no hair'' transition of the corresponding AdS black hole.

In the past few years, different classes of holographic superconductor have been explored \cite{G11,SCG0895,SCG0863,GM08,GM09,RLLR15,LWCLRL16,SS08,RSLL13,AY00,RLL14,FDJ11,AJ11,AJ12,JYDWC10,FCR10,LQJ15,LRLC15,RLLR14,EL20,YJWCJF14,YJCNXZS14,CQJY16,SQJ17,SWY17,CPD09,KAQE16,KAQE15,KEA15,KE14,MTWA09,S14,YQY17,QS16,QJBS14,romansan}, and significant progress has been made. 
For example, the holographic s-wave \cite{G11,SCG0895,SCG0863,GM08,GM09}, p-wave \cite{RLLR15,LWCLRL16,SS08,RSLL13,AY00,RLL14,FDJ11,AJ11,AJ12}, as well as d-wave superconductors \cite{JYDWC10,FCR10} have been investigated. 
As an s-wave superconductor is related to the instability and subsequent condensation of a charged scalar field, the p-wave and d-wave superconductors might be achieved by the condensations of a charged vector field and a spin-two field in bulk, respectively. 
The effect of the Weyl corrections on the Maxwell complex vector (MCV) superconductor in the AdS soliton and black hole \cite{LQJ15}, the holographic vector condensate and its related entanglement entropy \cite{RLLR15,LRLC15,RLLR14,EL20} and the superfluid \cite{YJWCJF14,YJCNXZS14,CQJY16,SQJ17,SWY17,CPD09} have been addressed by various authors. 
In particular, the p-wave superconductor can be realized by introducing a charged vector field in bulk as a vector order parameter, among others \cite{SS08,RSLL13,AY00,RLL14,FDJ11,AJ11,AJ12}.

The background geometry also plays a vital role in the study of the holographic superconductor. 
In addition to the schwarzschild AdS black hole geometry \cite{RSLL13}, the five-dimensional Gauss-Bonnet gravity with high curvature corrections has aroused much attention recently \cite{QJB11,SZ13,WJ13,XLR14,OL,ZH15,DS44,GY15,R02,RSB07,RSJ09,RZH10,QB10,HRH11,RZH11,QBEJA10,SD12}.
The problem in lower dimensions is constrained by the Coleman-Mermin-Wagner theorem which forbids spontaneous breaking of the continuous symmetry in the two- and three-dimensional space-times at finite temperature. 
Since holographic superconductors do exist in (2+1)-dimensions at finite temperature, such apparent contradiction is explained as the fluctuations in holographic superconductors are suppressed owing to the large $N$ limit
intrinsicly associated with the AdS/CFT correspondence. 
However, it is also known that higher curvature corrections suppress the condensation, it is therefore insteresting to investigate the properties of condensation in higher curvature gravity. 
In higher curvature case, for the theory to be ghost free one has to consider the appropriate combination of quadratic corrections, which naturally leads to the Gauss-Bonnet expression. 
The insulator/superconductor phase transition also has been investigated in the AdS soliton background \cite{TST10}, in the four-dimensional Einstein-Born-Infeld AdS theory by the MCV model beyond the probe limit \cite{PG15}, and the p-wave superfluid was studied in the MCV model for the AdS and the Lifshitz black hole metrics \cite{YJWCJF14,YJCNXZS14}. In a recent paper \cite{LWCLRL16}, the p-wave holographic superconductor was studied in the Gauss-Bonnet gravity. 
The authors examined the influences of Gauss-Bonnet parameter on the MCV model and found out that a large Gauss-Bonnet curvature correction hinders the phase transition, as the critical temperature decreases with increasing Gauss-Bonnet parameter.

The Lovelock theory of gravity \cite{CG08,RA13} is the most general metric theory of gravity built out of the fundamental assumptions of general relativity, which naturally restores to the Einstein's theory in three and four dimensions. 
By construction, the Gauss-Bonnet gravity can be viewed as a particular case of the Lovelock theory which remains invariant under the local Lorentz transformation. 
Therefore, it is a desirable gravity theory to describe the spacetime geometry in high dimensions. 
Recently, a particular Lovelock Lagrangian was proposed by Gaete and Hassa\"{i}ne where they consider a self-interacting scalar field nonminimally coupled to the gravity theory \cite{MM13}. 
As the coefficients of the Lovelock expansion in the model are determined by the requirement of a unique AdS vacuum with a fixed value of the cosmological constant, inequivalent gravity theories, indexed by an integer $k$, are obtained. 
For $k=1$, the theory restores to that of the standard Einstein-Hilbert. 
For dimension $d \geq 5$ and $k \geq 2$, two classes of planar AdS black hole solutions were obtained for the particular choices of the nonminimal coupling parameter $\xi$.
The authors show that the second class of solutions in $d$ dimension can be converted to the first class of solutions in $(d + 1)$ dimension through a Kaluza-Klein oxidation. 
These solutions are attributed to the existence of the higher order curvature terms of the underlying theory. 
The thermodynamics of the Lovelock black holes is also carried out in a later study for a generalized version of the model \cite{FM14}, where the authors found that an integration constant appearing in the black hole solution can be
interpreted as a sort of hair since there is not any conserved charge associated with it.

From the field theory point of view, higher curvature terms
correspond to $1/N$ corrections in an effective field theory.
Roughly speaking, these terms can be ignored only when the length
scale of the curvature is much larger than the string length. This
is usually satisfied for a strongly coupled system where 't Hooft
coupling is large. On the other hand, however, the inclusion of the
higher curvature contributions allows one to access the corrections
on the field theory side when the coupling is not too strong. In
this context, as Lovelock gravity is a more general theory with
higher curvature, its dual counterpart in field theory may reflect a
more realistic strongly coupled system with condensation. In
literature, one encounters essentially two different approaches for
holographic superconductors. The first approach, by and large, aims
at phenomenologically reproducing relevant features of the
high-temperature superconductor in terms of field operators by
writing down a bulk action with a minimum amount of degree of
freedom. While the second one is to construct the model from a
general theory on the gravity side while consistently including all
possible terms truncated at a given order. In the latter case, one
usually has a more complete and therefore better control over the
dual field theory. In this regard, by varying the parameter $k$, one
might be able to access a larger parameter space in the dual field
theory, and hopefully, some specific configuration might be more
realistic.

Because most known high-temperature superconductors are very
sophisticated compounds, it seems unlikely that a given holographic
setups will precisely capture the microscopic details of the
condensed matter system. Therefore, the philosophy is that the
holographic superconductor may uncover some universal aspects, such
as critical exponents and universality, of the strongly coupled
system, such as a high-temperature superconductor. As superconductor
transition is a second order phase transition, critical exponents
are interesting especially because they are observables closely
connected with measurements, while for the most part, independent of
specific details of the physical system. In turn, such approach
might improve our understanding of the underlying physics, in
particular, provide explicit examples of theories without the
picture of Fermi liquid. It was shown that critical exponents could
be extracted from the holographic superconductor~\cite{UC-01}. The
results agree with the Ginzburg-Landau theory, which is understood
since the Ginzburg-Landau theory is a mean-field theory, and the
holographic superconductor agrees with the results of the mean-field
theory because the former considers the large $N$ limit. In this
limit, quantum fluctuations are suppressed, so that mean-field
results are exact. According to the results of the renormalization
group, the critical exponents of a realistic physical system are
usually beyond those of the mean-field theory and belong to one of
the universality classes. In this context, since the inclusion of
the higher curvature contributions corresponds to high order
corrections on the field theory, it might provide more meaningful
physical content in terms of critical components. For the above
reasons, it is interesting to further investigate the p-wave
holographic superconductor in a black hole geometry of the Lovelock
gravity studied in \cite{MM13}.

In this paper, we employ the Maxwell-vector model to construct a
p-wave holographic superconductor in the background of the Lovelock
gravity. In bulk, the model introduces a complex vector field
nonminimally coupled to a local U(1) gauge field, the Maxwell field,
which is dual to a strongly coupled system concerning a charged
vector operator with a global U(1) symmetry on the boundary
\cite{RLLR15}. In the probe limit, we study numerically how the
superconductor phase transition is affected by the coupling
parameter, as well as the index $k$ of the particular Lovelock
theory. We show that the integer $k$, which is closely related to
the particular form of the nonminimal coupling parameter, indeed
plays a significant role in the superconductor phase transition.
Also, the dependence on the mass of the vector field and
dimensionality is investigated. The real, as well as the imaginary
parts of the conductivity of the holographic superconductor, is
analyzed and presented in function of $\omega/T_c$.

The rest of this paper is organized as follows. In Section II, we
construct the p-wave holographic superconductor model in the planar
AdS black hole geometry. The equations of motion and the asymptotic
solution at the boundary are derived. In Section III, we present the
numerical results by showing the condensate as a function of
temperature. The dependence of the results on $k$, the mass of the
vector field, and space-time dimension $d$ are also analyzed. In
Section IV, we investigate the conductivity as a function of the
ratio of the frequency to the critical temperature. Section V is
devoted to the discussions and conclusions.

\section{The p-wave holographic model}

For a particular Lovelock gravity which involves a self-interacting scalar field nonminimally coupled to the curved spacetimes, Gaete and Hassa\"{i}ne derived two classes of AdS black hole solutions.
These solutions are characterized by a planar event horizon topology and are obtained for the specific choices of the nonminimal coupling parameter $\xi$.
Here we study the vector condensation in the first family of planar AdS black hole metrics.
The black hole metric in $d\geq5$ dimension reads \cite{MM13}.
\begin{eqnarray}
ds^2 &=& -r^{2}f dt^2+\frac{dr^2}{r^{2}f} +r^{2}dx^2_{d-2} \, ,
\label{BH}\\
f &=& 1-\frac{M}{r^{\frac{d-2}{k}}} \, , \label{fexpr}
\end{eqnarray}
where the integer $k$ is to guarantee the theory to have a unique AdS vacuum with a fixed value of the cosmological constant.
This family of solutions is valid only for $k\geq2$.
It can be shown that the solution reduces to that in the Einstein-Gauss-Bonnet case for $k = 2$ \cite{MM13prd}.
The Hawking temperature of the black hole is
\begin{eqnarray}
T = \frac{(d-2)r_h}{4k\pi} \, ,
\end{eqnarray}
where $r_h=M^\frac{k}{d-2}$ is the radius of the black hole horizon.

To study the p-wave holographic superconductor, we consider the matter action by including a Maxwell field and a complex vector field \cite{RLLR15}
\begin{eqnarray}
\mathcal{I_{MCV}} &=& \frac{1}{16\pi G} \int d^dx \bigg\{\sqrt{-g}\Big[ -\frac{1}{4}F_{\mu\nu}F^{\mu\nu}-\frac{1}{2}(D_\mu\rho_\nu-D_\nu\rho_\mu)^\dagger(D^\mu\rho^\nu-D^\nu\rho^\mu) \nonumber \\
&& -m^2\rho^\dagger_\mu\rho^\mu+iq\gamma\rho_\mu\rho^\dagger_\nu
F^{\mu\nu}\Big] \bigg\} \, , \label{act}
\end{eqnarray}
where the Maxwell field strength tensor $F_{\mu\nu}=\nabla_\mu A_\nu-\nabla_\nu A_\mu$ and the covariant derivative $D_\mu=\nabla_\mu-iqA_\mu$.
The constants $q$ and $m$ correspond respectively to the charge and the mass of the vector field $\rho_\mu$.
The last term represents the nonminimal coupling between the vector field $\rho_\mu$ and the gauge field $A_\mu$, where $\gamma$ measures the magnetic moment of the vector field $\rho_\mu$.
In our study, we will neglect the magnetic field effects on the superconductor transition.

By variational principle, one obtains the following equations of the motion for the matter fields from the action Eq.(\ref{act})
\begin{eqnarray}
D^\nu(D_\nu\rho_\mu-D_\mu\rho_\nu)-m^2\rho_\mu = 0\, , \label{eq1} \\
\nabla^\nu
F_{\nu\mu}-iq[\rho^\nu(D_\nu\rho_\mu-D_\mu\rho_\nu)^\dagger-\rho^{\nu\dagger}(D_\nu\rho_\mu-D_\mu\rho_\nu)]
= 0 \, .\label{eq2}
\end{eqnarray}
We adopt the following ansatz for the vector field $\rho_\mu$
\begin{eqnarray}
\rho_\mu dx^\mu = \rho_x(r) dx
\end{eqnarray}
and \begin{eqnarray}
A =(\phi(r),~0,~0,~0,~\cdots)
\end{eqnarray}
for the gauge field $A_\nu$, where the dots indicate that the other components of the gauge field are null.
The above equations of motion can be expressed as,
\begin{eqnarray}
\rho_x''+ \left[\frac{f'}{f}+\frac{(d-2)}{r}\right]\rho_x'+\left[\frac{q^{2}\phi^2}{r^4f^2}-\frac{m^2}{r^2f}\right]\rho_x = 0 \, , \label{eom1} \\
\phi''+ \frac{(d-2)}{r}\phi'-\frac{2q^{2}\rho_x^2}{r^4f}\phi = 0 \,.
\ \label{eom2}
\end{eqnarray}

The boundary conditions at the horizon $r = r_h$ of the AdS$_d$ bulk are imposed by requiring that the vector field $\rho_\mu$ be regular, and the gauge field $A_\mu$ satisfy $\phi(r_h) =0$.
At the conformal boundary $r\rightarrow\infty$, the asymptotic forms of the matter field and the gauge field are given by
\begin{eqnarray}
\rho_x(r) = \frac{\rho_{x-}}{r^{\Delta_-}}+\frac{\rho_{x+}}{r^{\Delta_+}}+\cdots \, ,\label{asympsi} \\
\phi(r) = \mu-\frac{\rho}{r^{(d-3)}}+\cdots\, ,\label{asymphi}
\end{eqnarray}
where $\Delta_\pm = \frac{1}{2}[(d-3)\pm \sqrt{(d-3)^2+4m^2}]$, with the Breitenlohner-Freedman (BF) bound $m^2\geq -\frac{(d-3)^2}{4}$.
This implies that when $m^2= -\frac{(d-3)^2}{4}=m_{BF}^2$, $\Delta_+=\Delta_-=\Delta_{BF}=\frac{d-3}{2}$.
The coefficients $\mu, \rho$ and $\rho_{x-}, \rho_{x+}$ represent the chemical potential, charge density, the source and the $x$ component of the vacuum expectation value of the dual vector operator $<\mathcal{O}>$ respectively.
In addition, we will impose the condition $\rho_{x-} = 0$ to guarantee that the vector condensation will arise spontaneously in the boundary theory.

It is straightforward to verify that the equations of motion Eqs.(\ref{eom1}) and (\ref{eom2}) satisfy the following scaling law:
\begin{eqnarray}
r \to \lambda r ,\       (t.x,y,z) \to \frac{1}{\lambda}(t,x,y,z) \, ,\ \\
  \rho \to \lambda^{d-2} \rho ,\    (T,\mu) \to \lambda (T,\mu),\        \rho_{x+} \to \lambda^{1+\Delta_+}\rho_{x+} \, ,\
\end{eqnarray}
with a positive constant $\lambda$.
In what follows, we solve numerically the above nonlinear differential equations Eqs.(\ref{eom1}) and (\ref{eom2}) by using the shooting method \cite{SCG0895,SCG0863}.
We use the scaling symmetries to set $r_h = 1$ and $q=1$, so that the solution of the above equations can be expressed in terms of two independent parameters defined at the horizon, namely, $\phi'(r_h)$ and $\rho_x(r_h)$.
We use one of them to ensure the source free condition, $\rho_{x-} = 0$.
The rest physical quantities, such as $\mu,~\rho,~\rho_{x+}$, can be obtained by reading off the corresponding coefficients in the asymptotic forms as given in Eqs.(\ref{asympsi}) and (\ref{asymphi}).
To study the temperature dependence of the condensate, we make use of the scaling symmetry mentioned above and always express our result regarding scale invariant quantities, such as $T_c/\rho^{\frac{1}{d-2}}$ and $<\mathcal{O}>^{\frac{1}{1+\Delta_+}}/T_c$.
By doing this, the results obtained by varying the charge density $\rho$ for a given $r_h$ can be interpreted equivalently as those obtained for given $\rho=1$ and variant temperature.

\section{Condensation of the vector field}

Now we present the numerical results on the condensate of the massive vector field for different values of $m^2=-3/4, 0, 5/4$ ($\Delta_+=\Delta=3/2, 2, 5/2$), in the $d=5$ dimensional black hole background, and those for $m^2=0$ ($\Delta_+=\Delta=3$) in the $d=6$ dimensional case.
The dependence on the index $k$ of the Lovelock theory is also studied.

\begin{table}[ht]
\caption{The calculated critical temperatures for different values
of vector field mass $m$ and index $k$.}
\begin{tabular}{c|c|c|c|c}
\hline
&  $d=5,m^2=-3/4$ &  $d=5,m^2=0$&  $d=5,m^2=5/4$&  $d=6,m^2=0$\\
\hline
k=2&  0.09646&    0.08497&    0.07733&    0.11847\\
\hline
k=3&    0.07280&    0.06185&    0.05557&    0.08426\\
\hline
k=4&    0.06230&    0.04987&    0.04419&    0.06668\\
\hline
k=5&    0.05902&    0.04238&    0.03706&    0.05580\\
\hline \label{tb1}
\end{tabular}
\end{table}

Tab.\ref{tb1} shows the calculated critical temperature for different parameters.
The critical temperature $T_c$ is inverse proportional to $\rho^{1/3}$ in $d=5$ and $\rho^{1/4}$ in $d=6$.
For a fixed value of $k$, the ratio $T_c/\rho^{1/3}$ decreases as $\Delta$ or $m^2$ increases.
Hence it becomes more difficult for the vector field to condense as its mass increases.
Moreover, the critical temperature decreases with increasing $k$, which implies
that increasing the value of integer $k$ hinders the conductor/superconductor phase transition.
It is also observed that the magnitude of the critical temperature decreases for nearly 100\% as $k$ increases from $2$ to $5$.
Furthermore, for given a vector mass ($m^2 = 0$) and an index $k$, the critical temperatures in the six-dimensional case are higher than those in the five-dimensional case.

\begin{figure}[ht]
\begin{center}

\begin{minipage}[c]{0.5\textwidth}
\centering\includegraphics[width=0.9\textwidth]{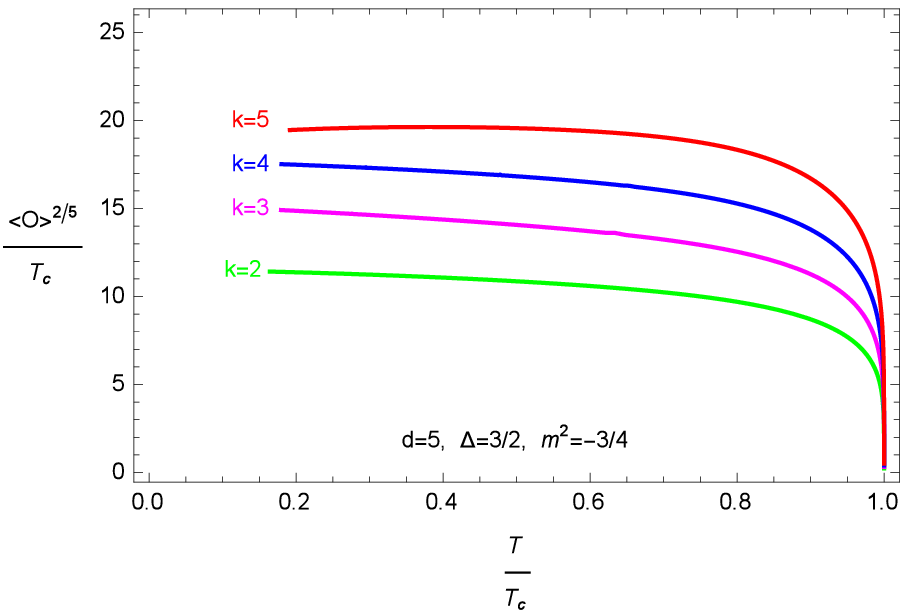}
\end{minipage}%
\begin{minipage}[c]{0.5\textwidth}
\centering\includegraphics[width=0.9\textwidth]{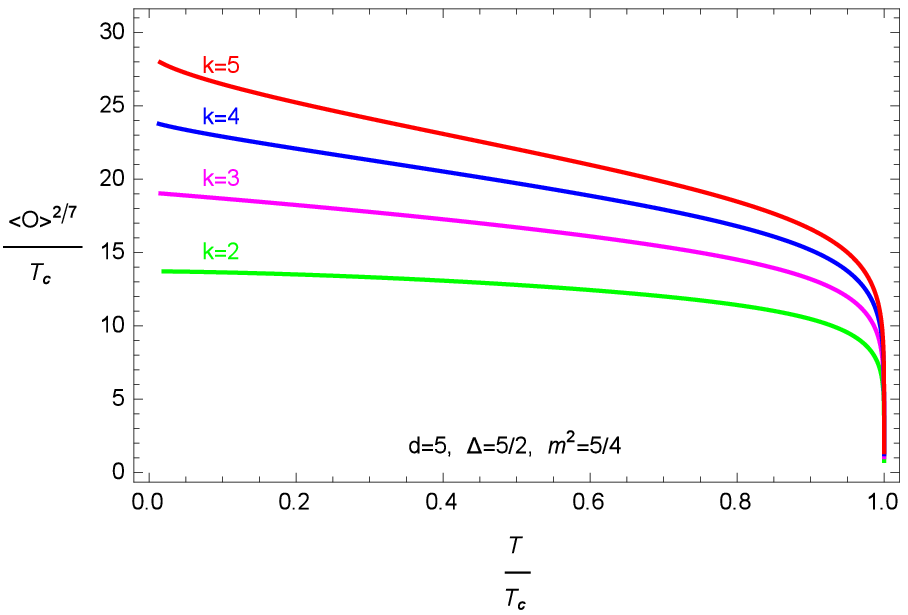}
\end{minipage}
\\
\begin{minipage}[c]{0.5\textwidth}
\centering\includegraphics[width=0.9\textwidth]{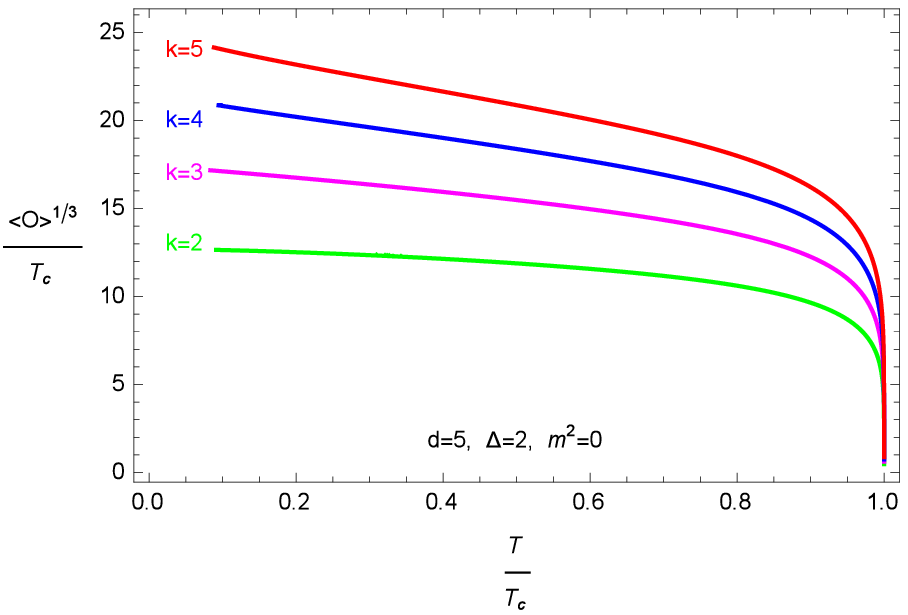}
\end{minipage}%
\begin{minipage}[c]{0.5\textwidth}
\centering\includegraphics[width=0.9\textwidth]{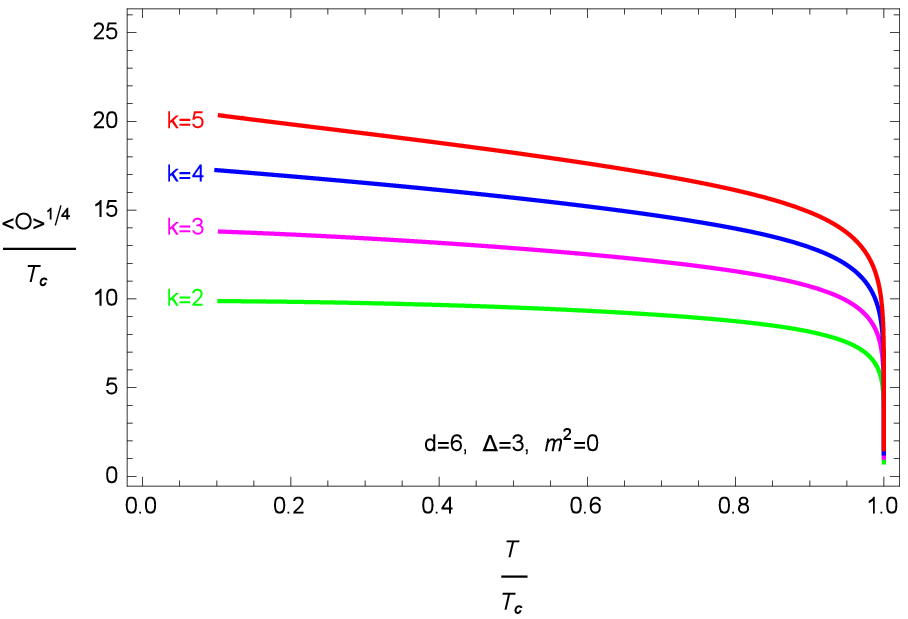}
\end{minipage}

\renewcommand{\figurename}{Fig.}
\caption{(Color online) The calculated vector condensate as a
function of temperature for different index $k$ with the fixed
vector field mass $m$ and dimension $d$.} \label{condensate1}
\end{center}
\end{figure}

In Fig.\ref{condensate1}, we show the calculated condensate as a
function of the temperature for different index $k$ with the fixed
vector field mass $m$ and dimension $d$. Firstly, to observe the
effects of the index $k$, the calculations are carried out with
$k=2, 3, 4$ and $5$. It is found that the vector condensation gap
becomes larger as $k$ increases. In other words, a bigger index $k$
makes it harder for the vector field to condense, and therefore more
difficult for the system to transition to the superconductor phase.
So the index $k$ labeling the particular Lovelock action is not only
crucial for the theory to have a unique AdS vacuum, but also has a
sizable impact on the vector condensation. Secondly, we investigate
the effect of the vector field mass $m$ by repeating the
calculations for different masses of the vector field $m^2=-3/4, 0,
5/4$ (with $\Delta_+=\Delta=3/2, 2, 5/2$ respectively) in the
five-dimensional case. One observes that as $m^2$ increases (and
therefore $\Delta$ increases), the vector condensation gap becomes
larger for a given $k$. Meanwhile, in all these cases, the
condensate with $k=2$ is the least sensitive to the temperature
before the critical temperature is reached. In fact, the condensate
in question is almost a constant, and its magnitude is more than
twice of that obtained in the weak coupling BCS theory, which is
$3.5$ \cite{JLJ57}(what is the value of condensate in BCS theory).
This confirms that the present p-wave model is of strongly coupled
field theory. However, the condensation curve for $\Delta=5/2$ and
$k=5$ appears to diverge at low temperature since we neglect the
backreaction of the spacetime. Lastly, we also do computations for a
massless vector field in the six-dimensional spacetime and find that
the vector condensation gaps in the six-dimensional case are smaller
than those in the five-dimensional case for a given $k$. This
implies that it is easier for vector condensation to take place in
the higher dimensional spacetime.

\begin{figure}[ht]
\begin{center}
\begin{minipage}[c]{0.5\textwidth}
\centering\includegraphics[width=0.9\textwidth]{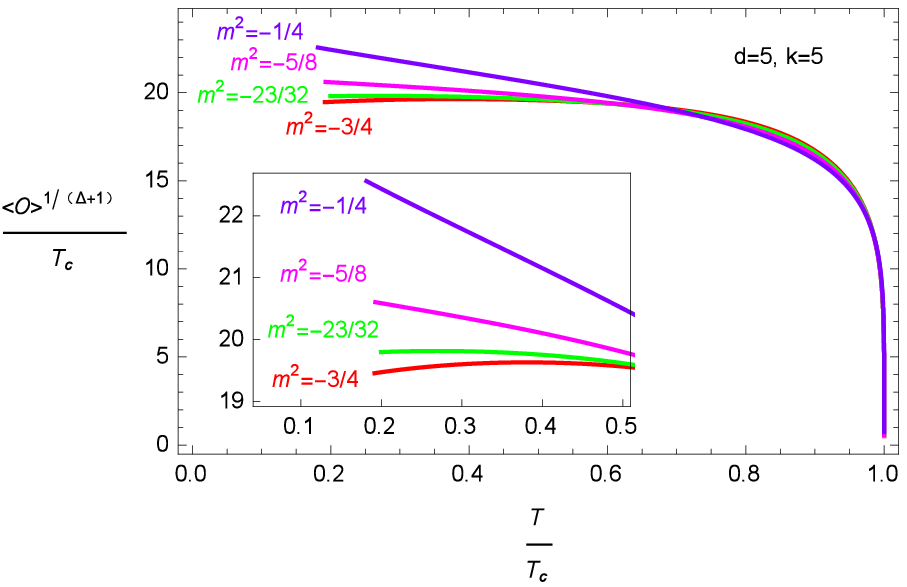}
\end{minipage}%
\begin{minipage}[c]{0.5\textwidth}
\centering\includegraphics[width=0.9\textwidth]{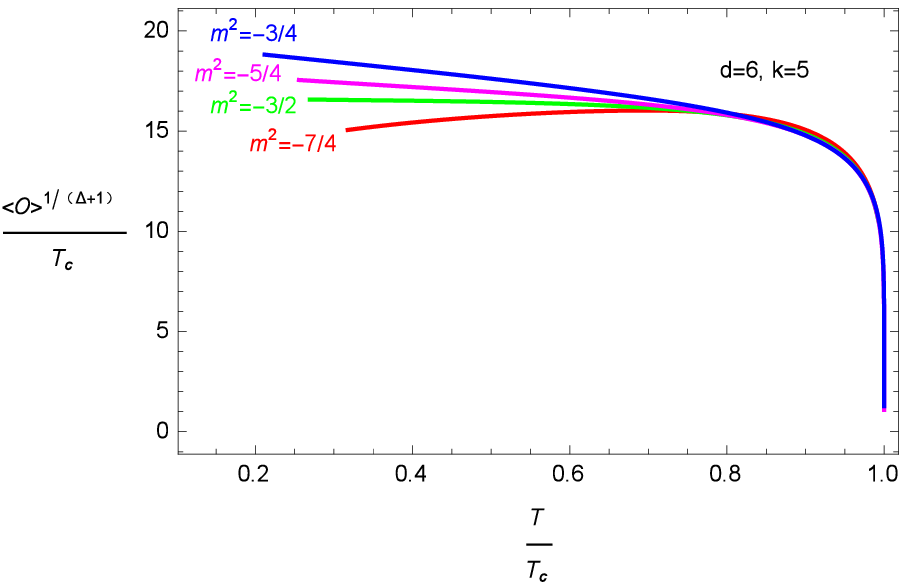}
\end{minipage}
\renewcommand{\figurename}{Fig.}
\caption{(Color online) The calculated vector condensate as a
function of temperature for different vector field mass $m$ with the
fixed index $k$ and dimension $d$.} \label{condensate2}
\end{center}
\end{figure}

In order to further study the validity of the probe approximation that has been used in our calculation, in Fig.\ref{condensate2} we
show the condensation curves for different $m^2$ with the fixed
index $k=5$ and dimensions $d=5$ and $6$ respectively. It is found
that the diverging behavior discussed above is related both to the
value of $k$ and $m^2$. For a given $k$ and $d$, as $m^2$ increases,
the condensate becomes divergent at small temperature $T/T_c$ as the
probe limit breaks down. For the case of $d=5$ and $k=5$, the
obvious divergence takes place at $m^2 \sim -\frac{1}{4}$, and for
the case of $d=6$ and $k=5$, it takes place at $m^2 \sim
-\frac{3}{4}$. There exists a translating value of $m^2$ beyond
which a diverging behavior appears, i.e., $m^2 \sim -0.72$ for the
case of $d=5$ and $k=5$ and $m^2 \sim -1.50$ for the case of $d=6$
and $k=5$. Considering the findings of Fig.\ref{condensate1}, we
observe that as the background changes, in terms of the index $k$,
the range of valid value of $m^2$ also modifies. However, it should
be noted that how the parameters $k$ and $m^2$ works is still an
open question.

\section{Conductivity}

In this Section, we study the conductivity \cite{SIA98,SCG0895,SS08,G11,RSLL13,RLLR15,PG15} by considering the perturbed Maxwell field $\delta A_y=A_y(r)e^{-i\omega t}dy$ in the case of the five-dimensional AdS black hole background.
The Maxwell equation reads
\begin{eqnarray}
A_y''+\left(\frac{f'}{f}+\frac{3}{r}\right)A_y'+\left(\frac{\omega^2}{r^4f^2}-\frac{2\rho_{x}^2}{r^4f}\right)A_y=0\,
,\label{eqAy}
\end{eqnarray}
with the ingoing wave boundary conditions at the horizon $(r = r_h =
1)$
\begin{eqnarray}
A_y(r)\sim (r-1)^{\frac{-i\omega k}{3}},
\end{eqnarray}
and the asymptotic behavior at infinity $r\rightarrow\infty$
\begin{eqnarray}
A_y(r)=A^{(0)}+\frac{A^{(2)}}{r^2}+\frac{A^{(0)}\omega^2}{2}\frac{\log\Lambda
r}{r^2}+\cdots.
\end{eqnarray}
It can be shown that the conductivity is given by
\begin{eqnarray}
\sigma(\omega)=\frac{G^R}{i\omega}=-\frac{2i A^{(2)}}{\omega
A^{(0)}}+\frac{i\omega}{2}
\end{eqnarray}
where the Green function satisfies
\begin{eqnarray}
G^R=-\lim_{r\to \infty} {r^3 f\frac{A'_y(r)}{A_y(r)}}\;.\nonumber
\end{eqnarray}

\begin{figure}[ht]
\begin{center}

\begin{minipage}[c]{0.3\textwidth}
\centering\includegraphics[width=1\textwidth]{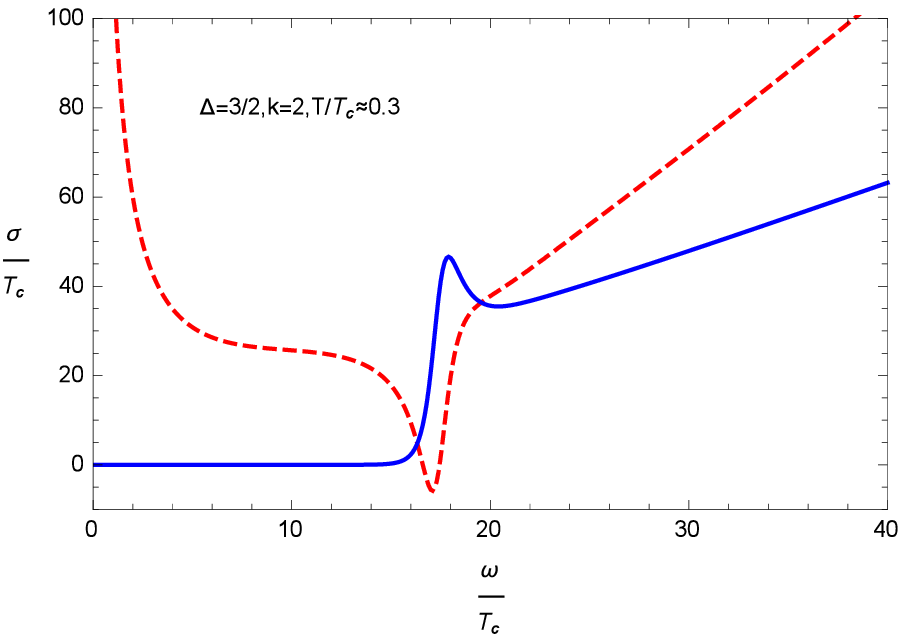}
\end{minipage}%
\begin{minipage}[c]{0.3\textwidth}
\centering\includegraphics[width=1\textwidth]{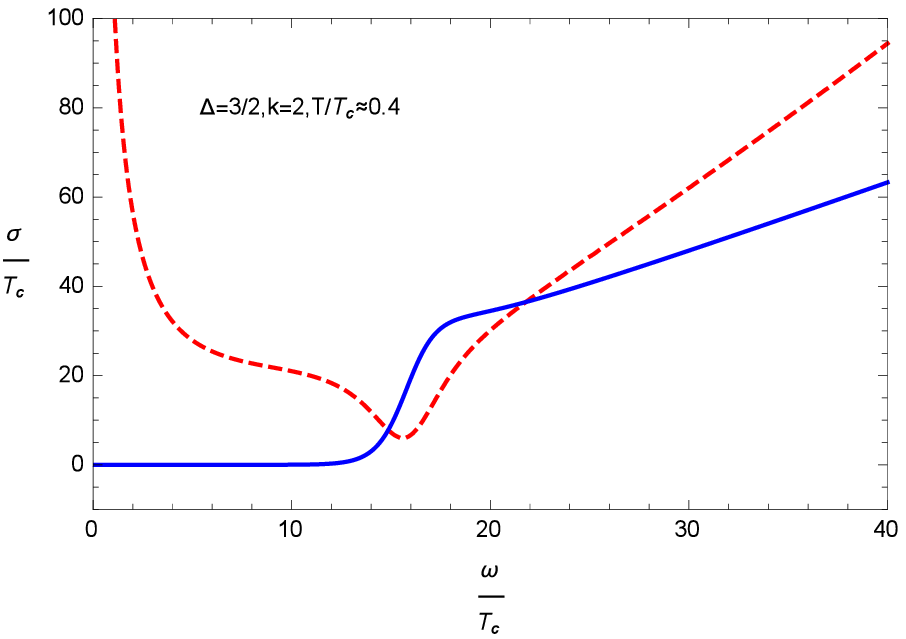}
\end{minipage}%
\begin{minipage}[c]{0.3\textwidth}
\centering\includegraphics[width=1\textwidth]{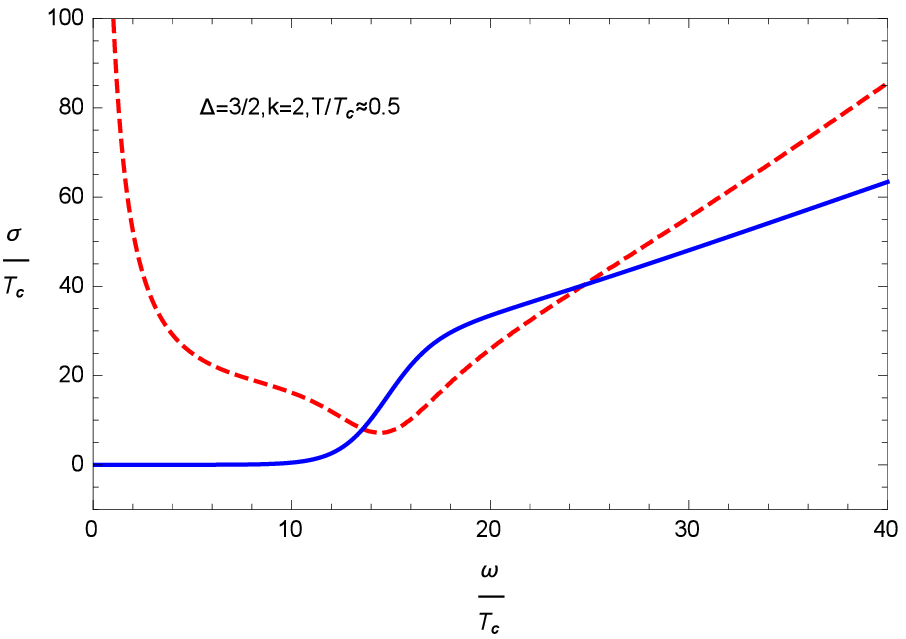}
\end{minipage}
\\
\begin{minipage}[c]{0.3\textwidth}
\centering\includegraphics[width=1\textwidth]{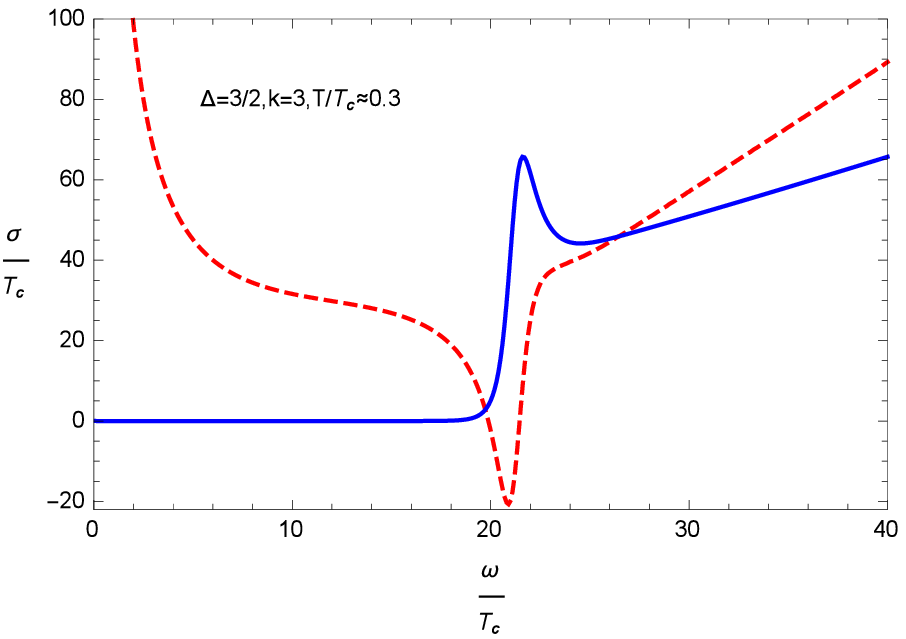}
\end{minipage}%
\begin{minipage}[c]{0.3\textwidth}
\centering\includegraphics[width=1\textwidth]{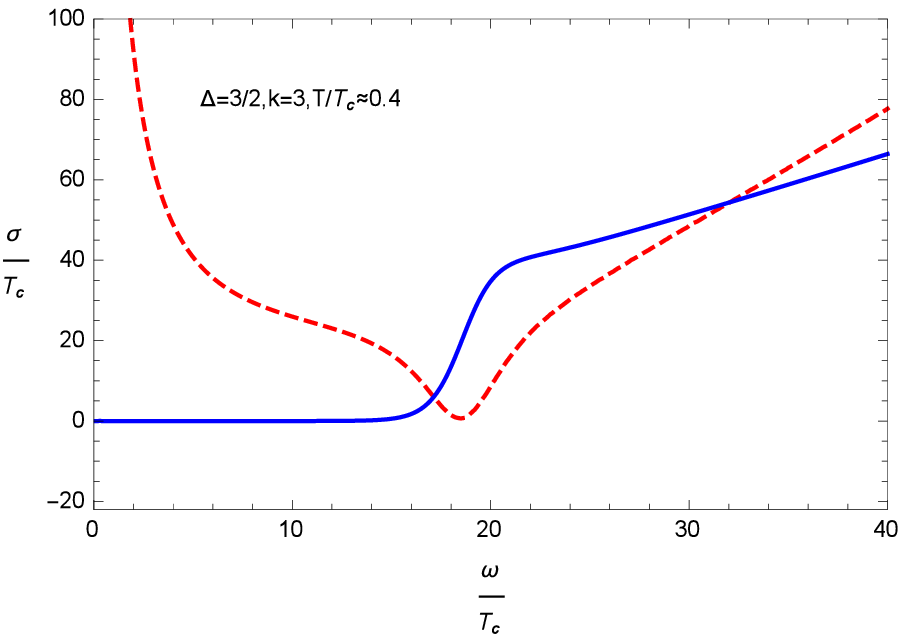}
\end{minipage}%
\begin{minipage}[c]{0.3\textwidth}
\centering\includegraphics[width=1\textwidth]{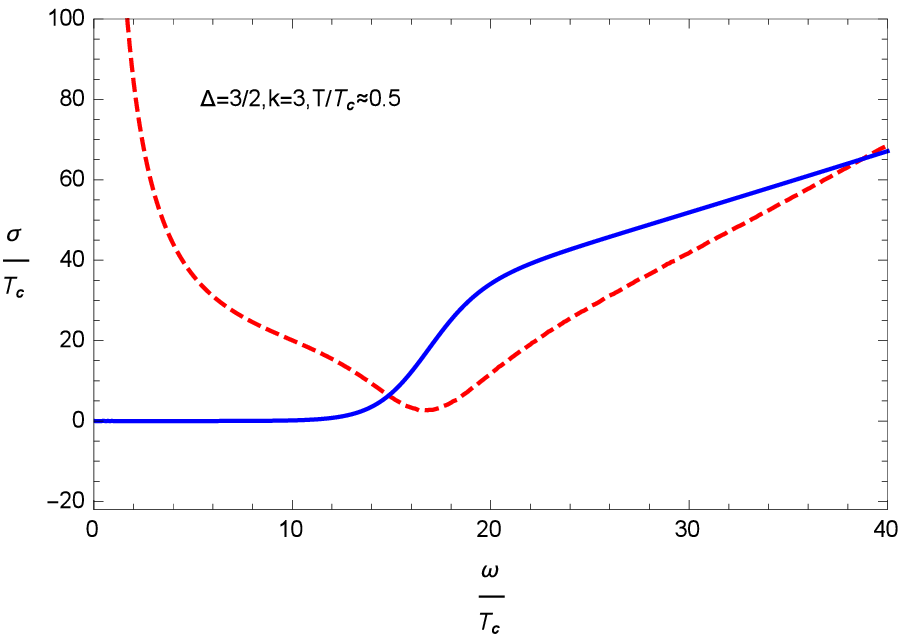}
\end{minipage}
\\
\begin{minipage}[c]{0.3\textwidth}
\centering\includegraphics[width=1\textwidth]{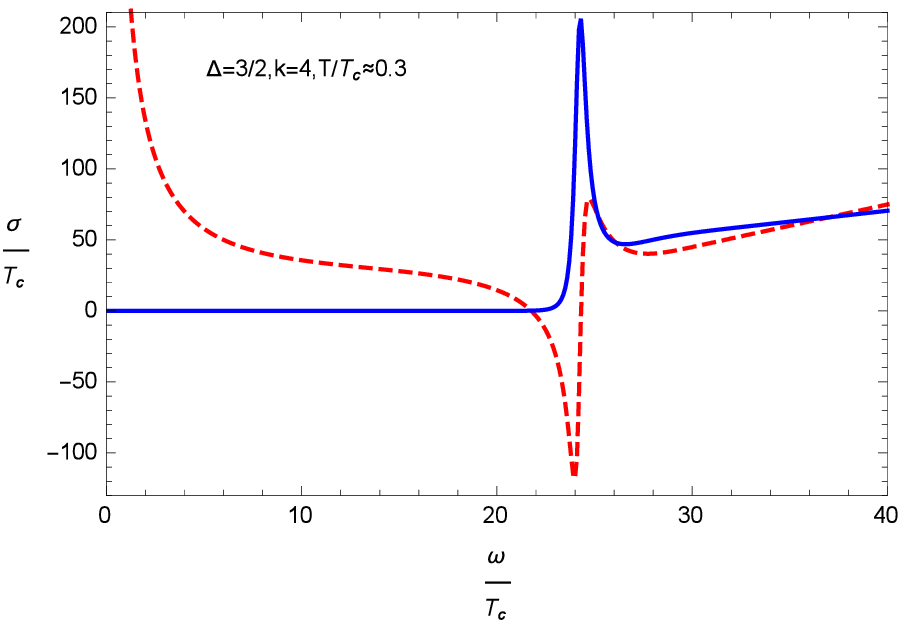}
\end{minipage}%
\begin{minipage}[c]{0.3\textwidth}
\centering\includegraphics[width=1\textwidth]{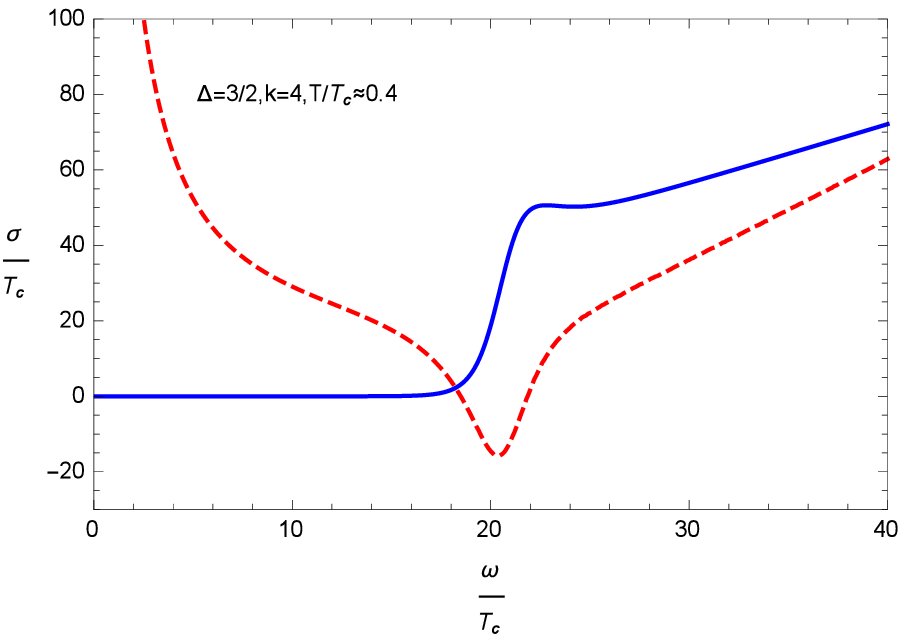}
\end{minipage}%
\begin{minipage}[c]{0.3\textwidth}
\centering\includegraphics[width=1\textwidth]{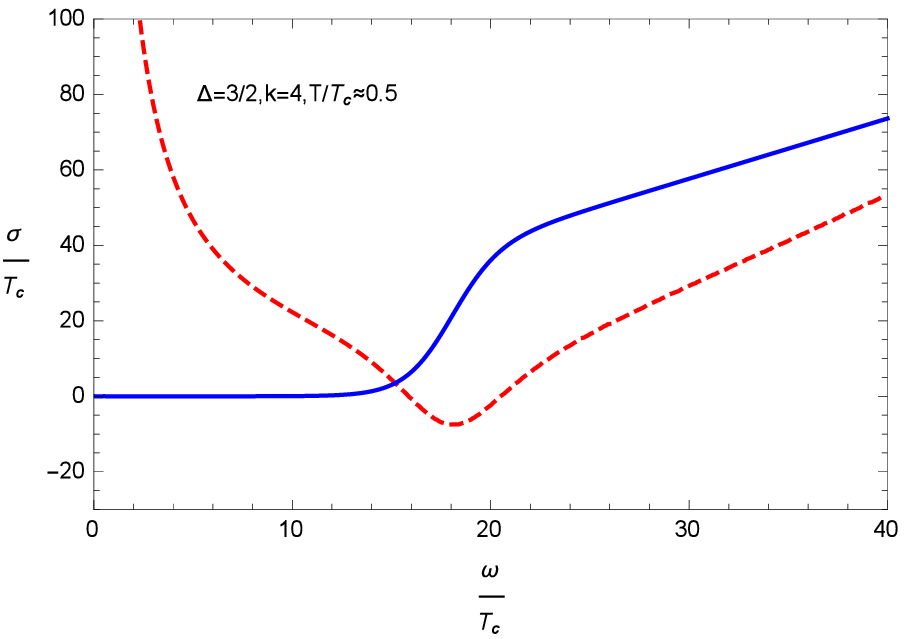}
\end{minipage}
\\
\begin{minipage}[c]{0.3\textwidth}
\centering\includegraphics[width=1\textwidth]{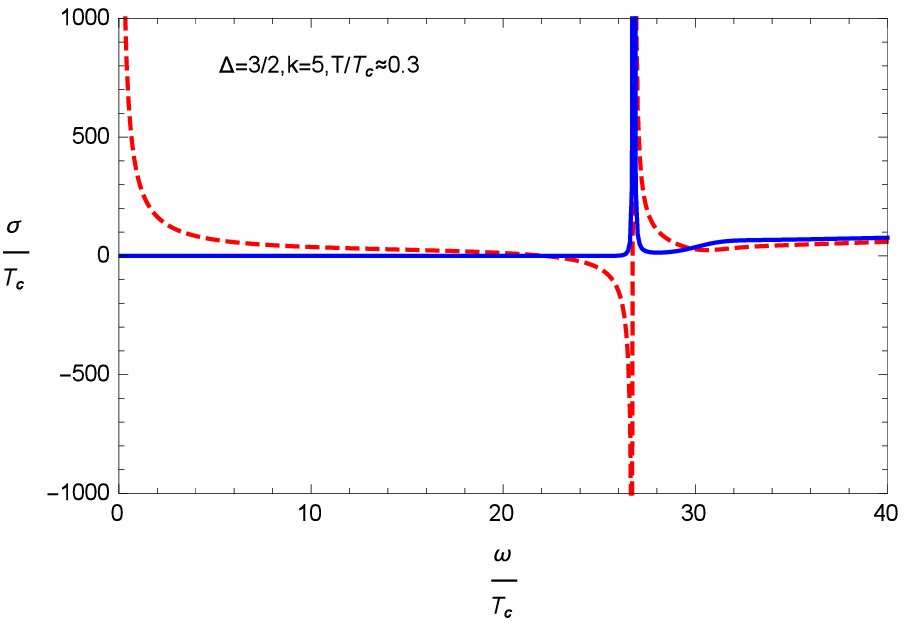}
\end{minipage}%
\begin{minipage}[c]{0.3\textwidth}
\centering\includegraphics[width=1\textwidth]{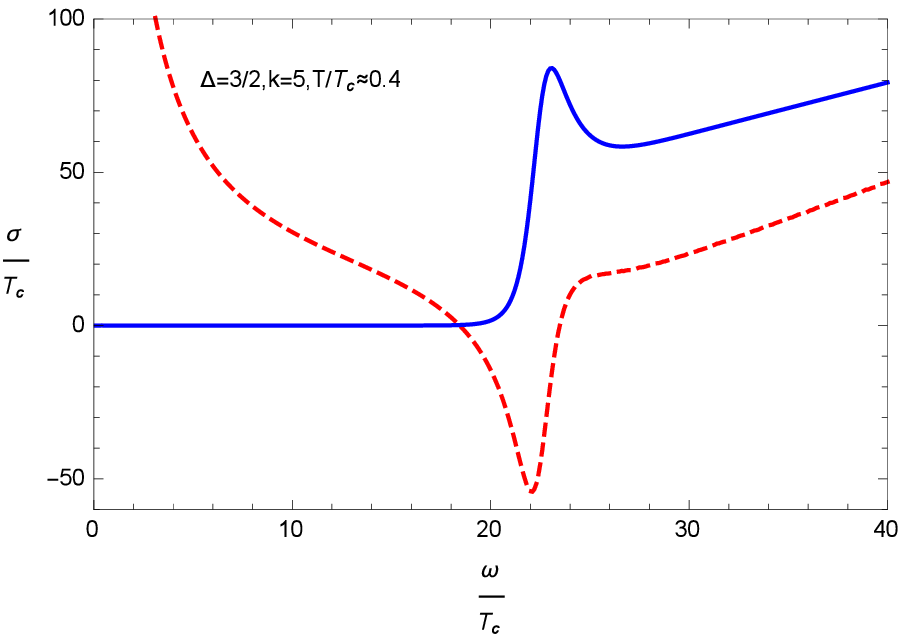}
\end{minipage}%
\begin{minipage}[c]{0.3\textwidth}
\centering\includegraphics[width=1\textwidth]{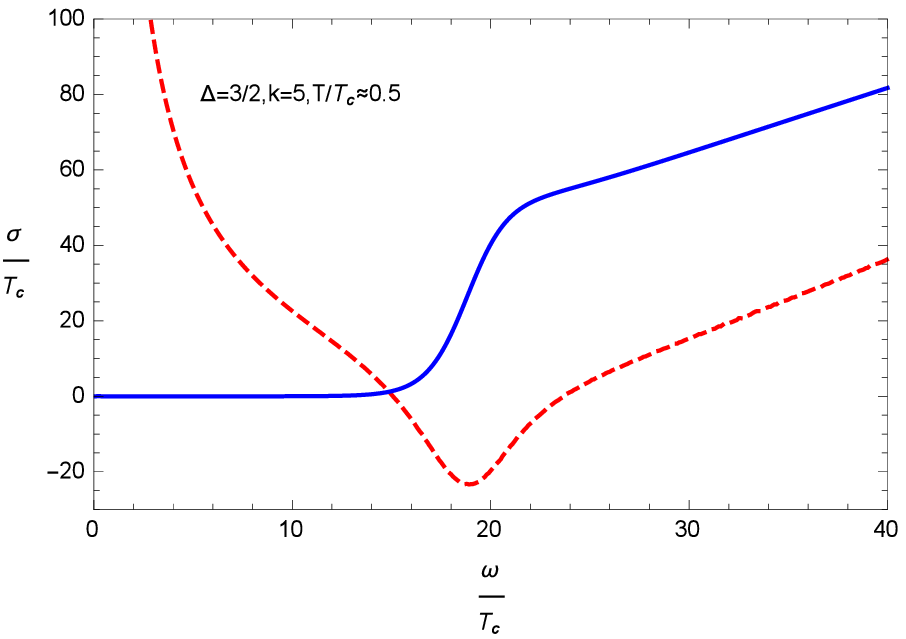}
\end{minipage}

\renewcommand{\figurename}{Fig.}
\caption{(Color online) The calculated conductivity of the holographic superconductor. The real and the imaginary parts of the conductivity are shown in solid blue and dashed red curves. The calculations are carried out for $\Delta=3/2, m^2=-3/4$.}
\label{conductivity1}
\end{center}
\end{figure}

\begin{figure}[ht]
\begin{center}

\begin{minipage}[c]{0.3\textwidth}
\centering\includegraphics[width=1\textwidth]{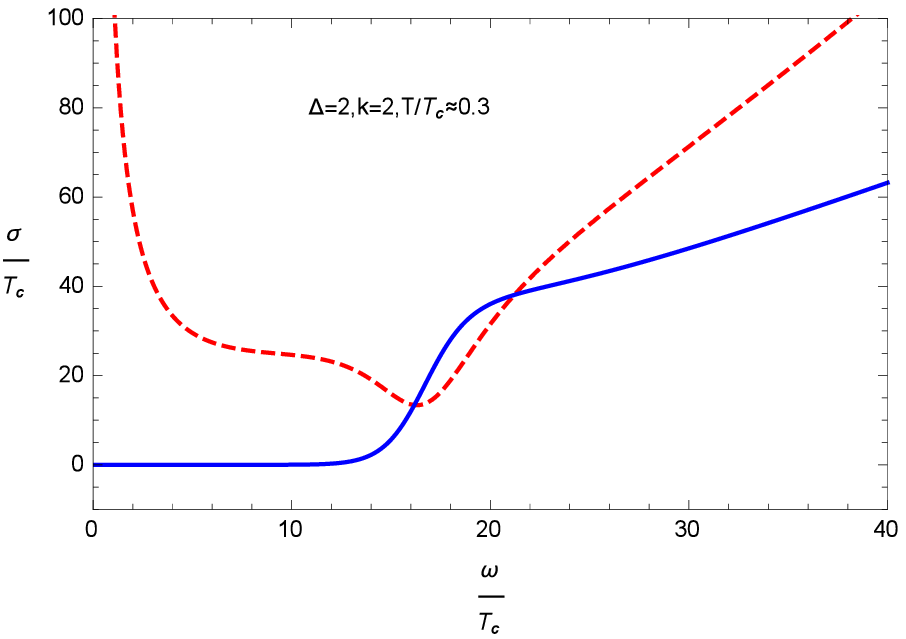}
\end{minipage}%
\begin{minipage}[c]{0.3\textwidth}
\centering\includegraphics[width=1\textwidth]{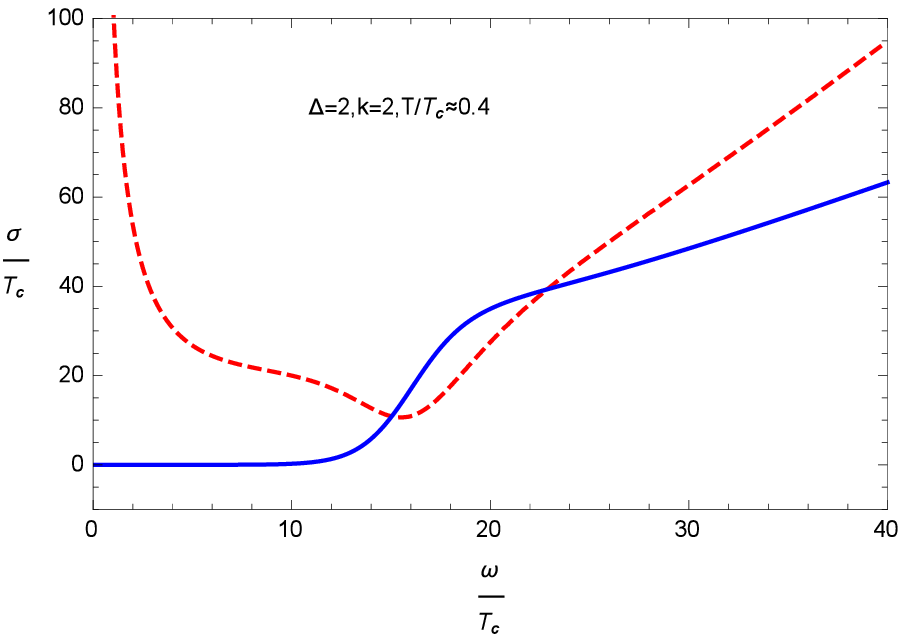}
\end{minipage}%
\begin{minipage}[c]{0.3\textwidth}
\centering\includegraphics[width=1\textwidth]{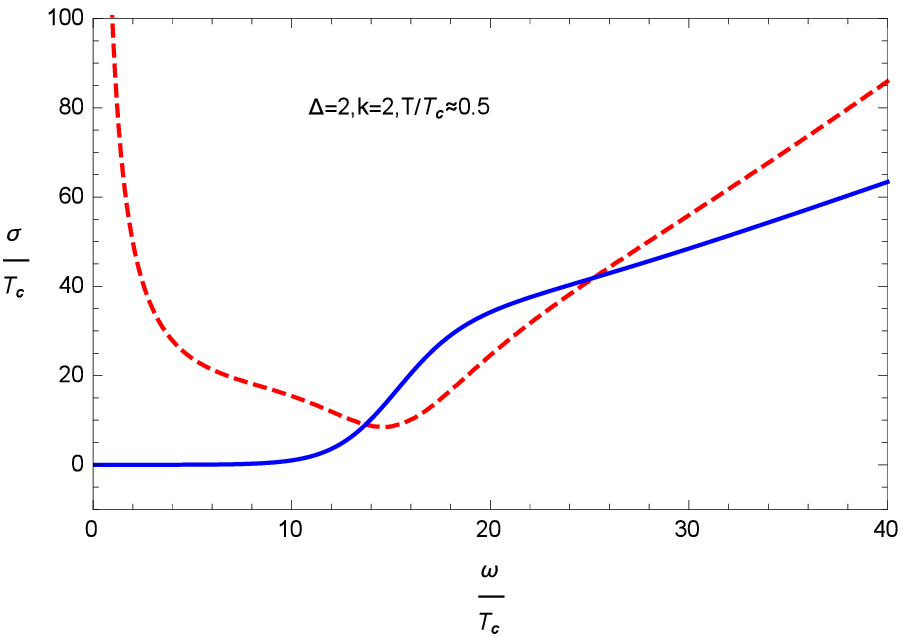}
\end{minipage}
\\
\begin{minipage}[c]{0.3\textwidth}
\centering\includegraphics[width=1\textwidth]{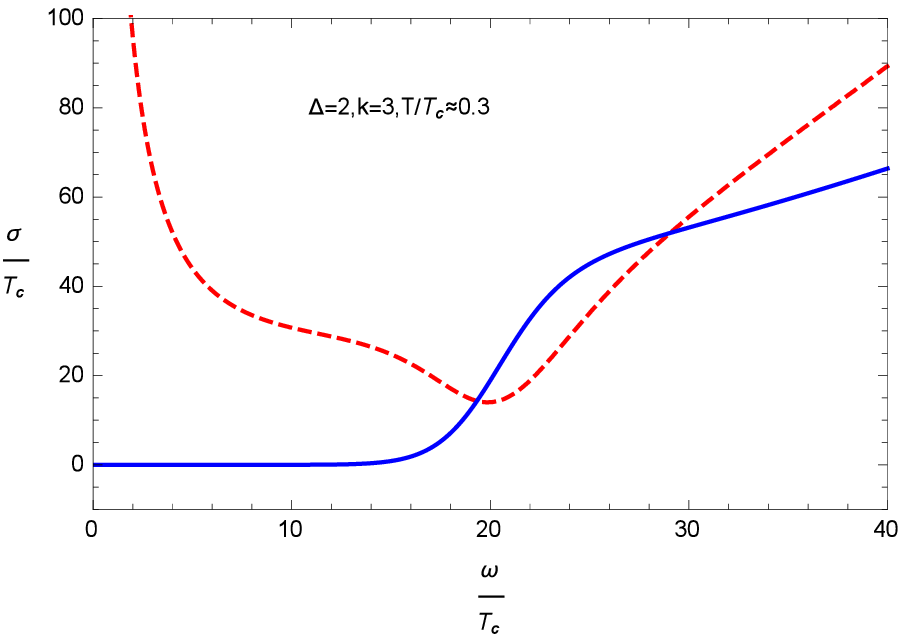}
\end{minipage}%
\begin{minipage}[c]{0.3\textwidth}
\centering\includegraphics[width=1\textwidth]{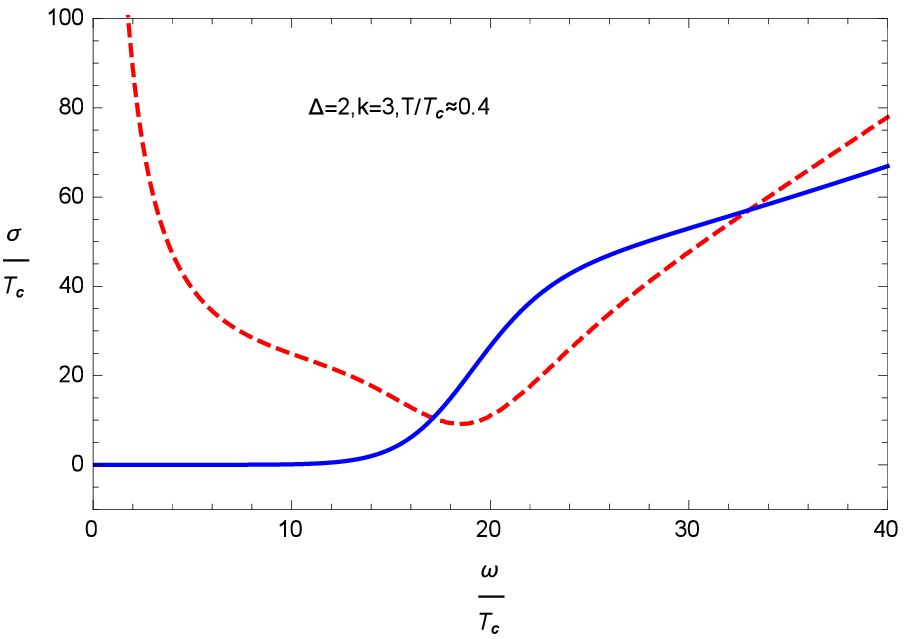}
\end{minipage}%
\begin{minipage}[c]{0.3\textwidth}
\centering\includegraphics[width=1\textwidth]{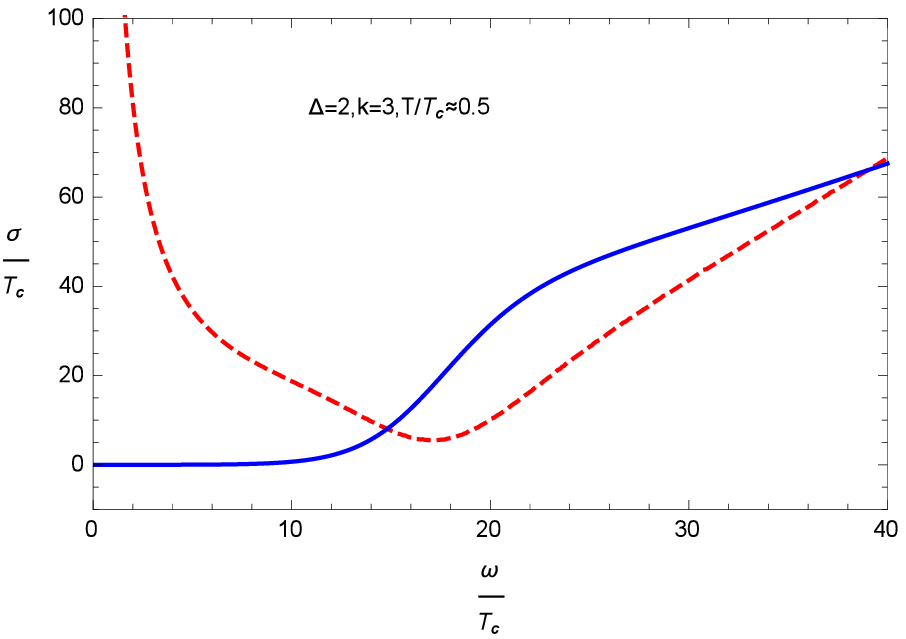}
\end{minipage}
\\
\begin{minipage}[c]{0.3\textwidth}
\centering\includegraphics[width=1\textwidth]{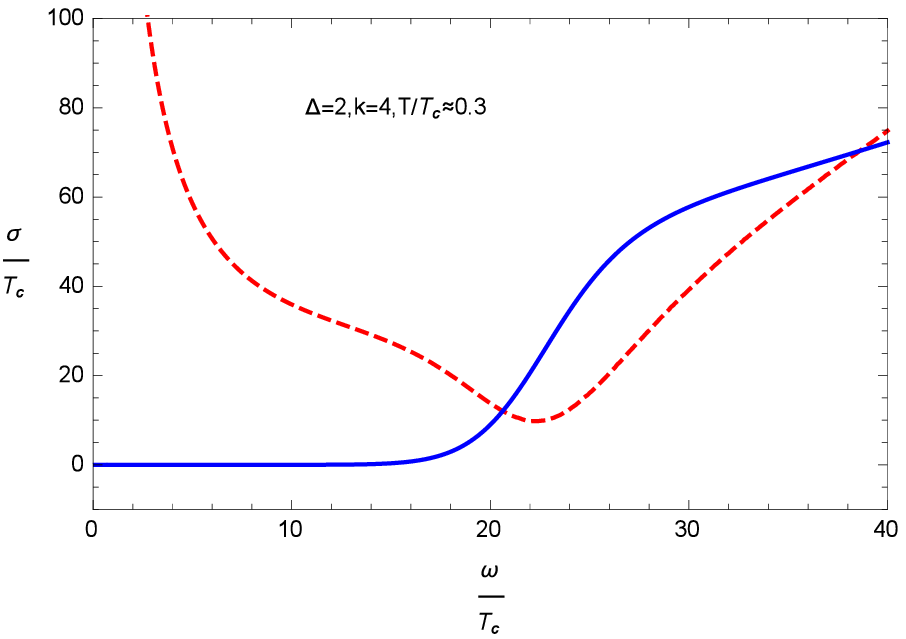}
\end{minipage}%
\begin{minipage}[c]{0.3\textwidth}
\centering\includegraphics[width=1\textwidth]{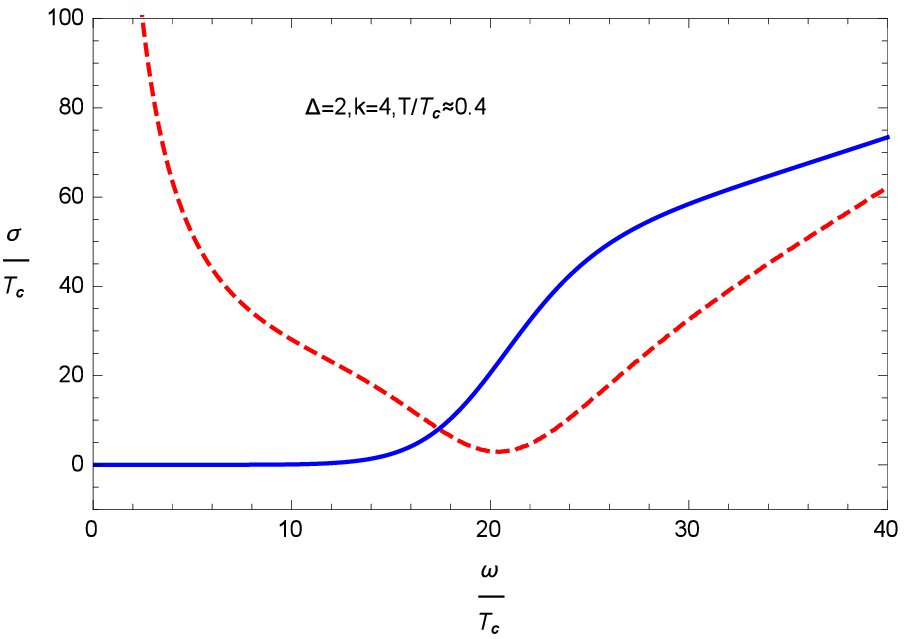}
\end{minipage}%
\begin{minipage}[c]{0.3\textwidth}
\centering\includegraphics[width=1\textwidth]{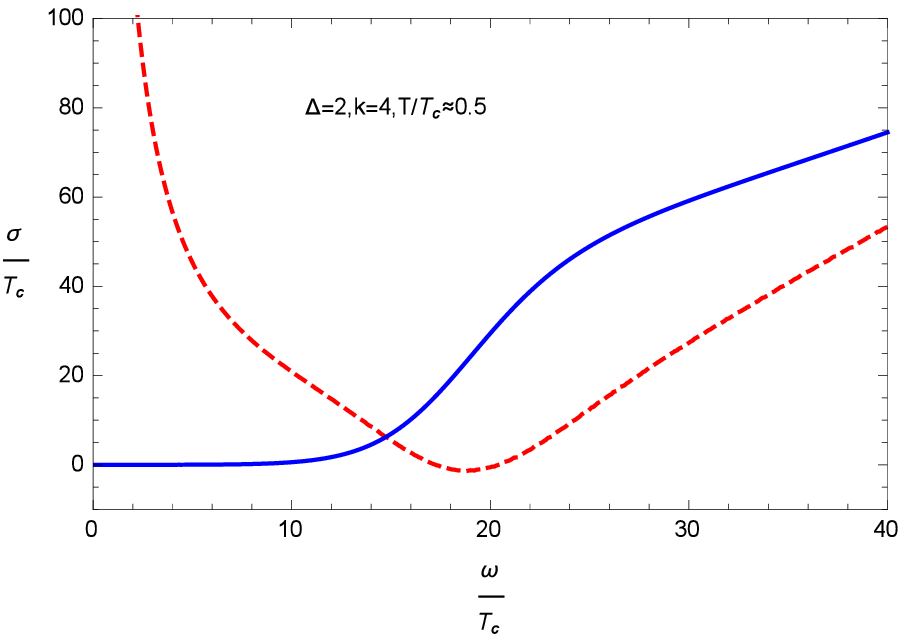}
\end{minipage}
\\
\begin{minipage}[c]{0.3\textwidth}
\centering\includegraphics[width=1\textwidth]{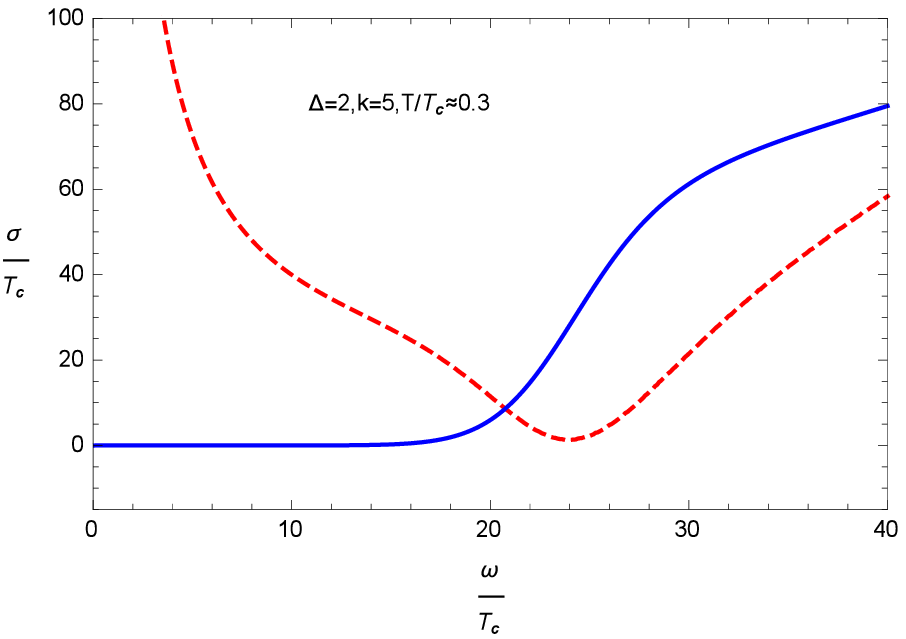}
\end{minipage}%
\begin{minipage}[c]{0.3\textwidth}
\centering\includegraphics[width=1\textwidth]{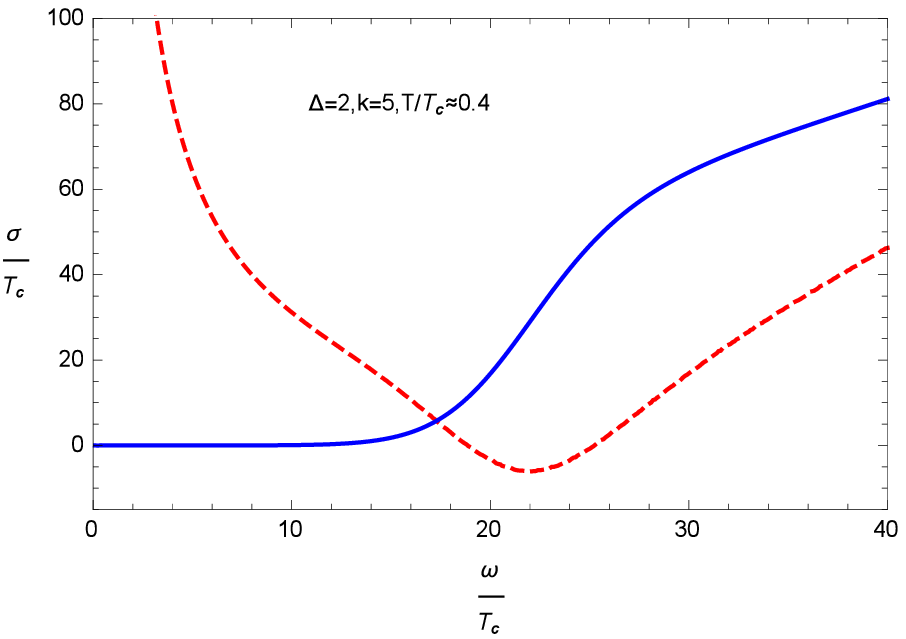}
\end{minipage}%
\begin{minipage}[c]{0.3\textwidth}
\centering\includegraphics[width=1\textwidth]{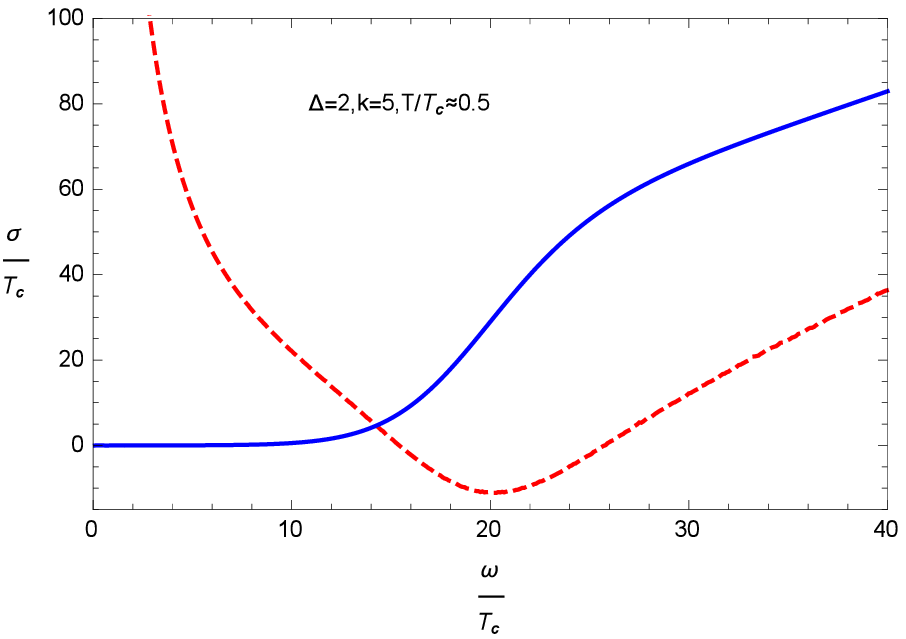}
\end{minipage}

\renewcommand{\figurename}{Fig.}
\caption{(Color online) The calculated conductivity of the holographic superconductor.
The real and the imaginary parts of the conductivity are shown in solid blue and dashed red curves.
The calculations are carried out for $\Delta=2, m^2=0$.}
\label{conductivity2}
\end{center}
\end{figure}

\begin{figure}[ht]
\begin{center}

\begin{minipage}[c]{0.3\textwidth}
\centering\includegraphics[width=1\textwidth]{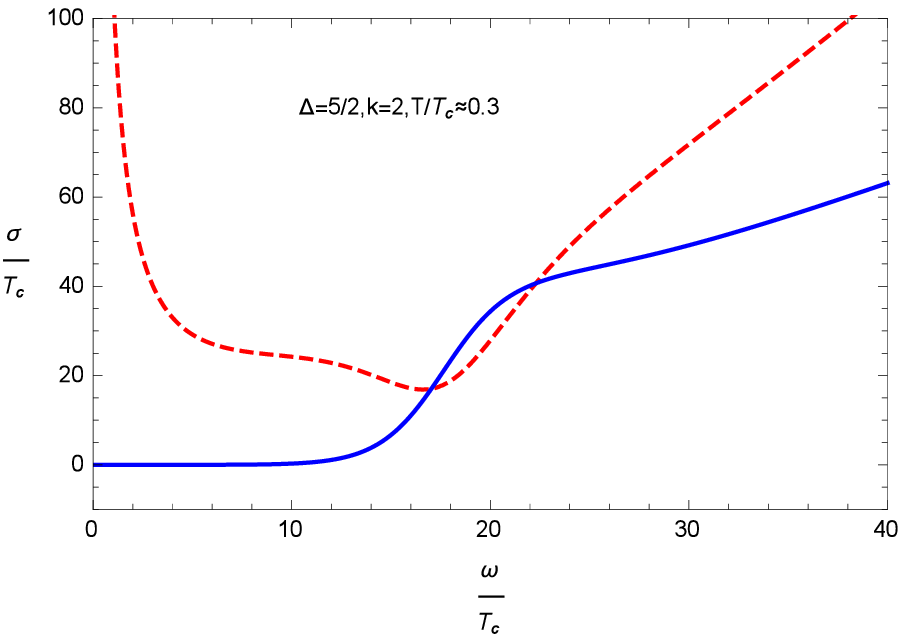}
\end{minipage}%
\begin{minipage}[c]{0.3\textwidth}
\centering\includegraphics[width=1\textwidth]{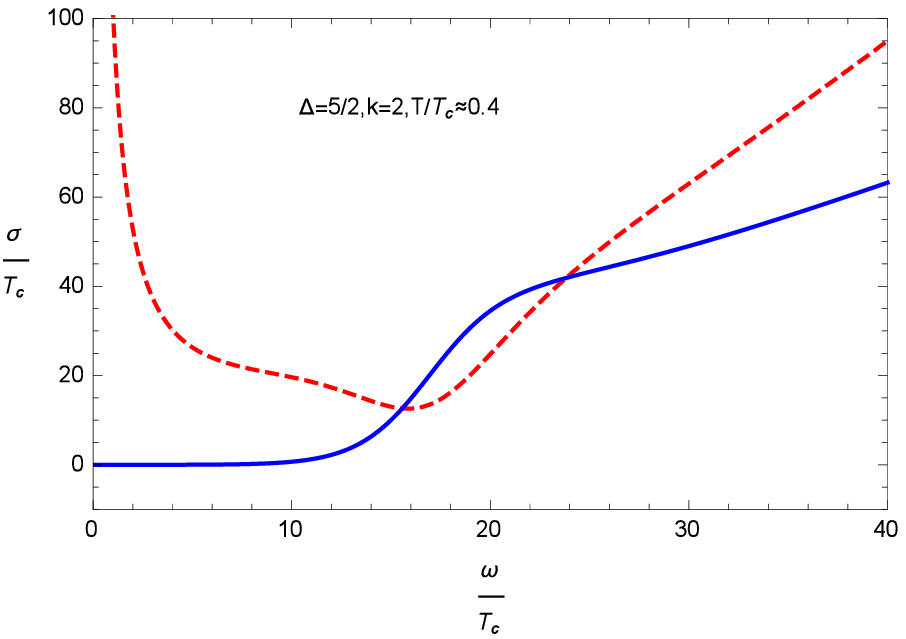}
\end{minipage}%
\begin{minipage}[c]{0.3\textwidth}
\centering\includegraphics[width=1\textwidth]{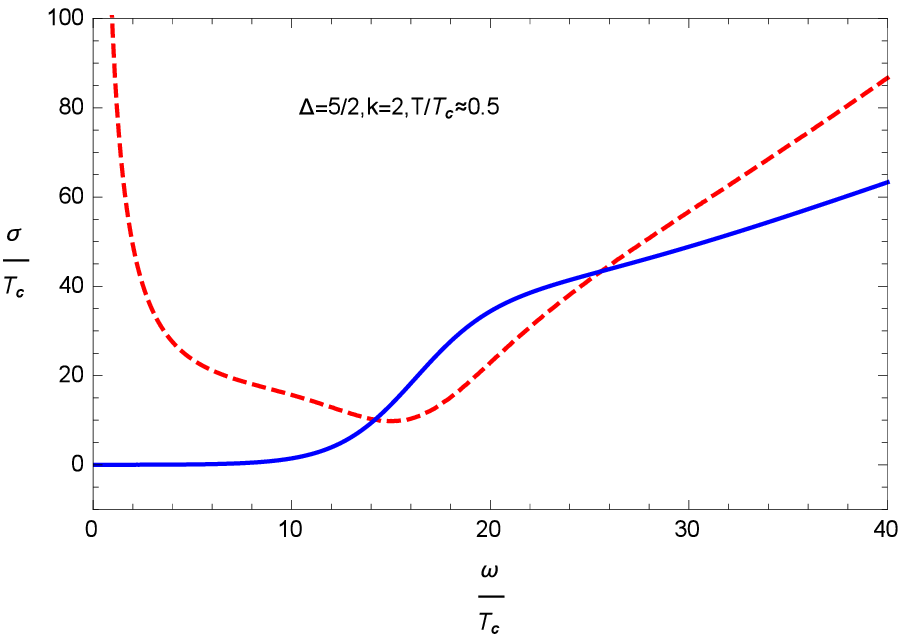}
\end{minipage}
\\
\begin{minipage}[c]{0.3\textwidth}
\centering\includegraphics[width=1\textwidth]{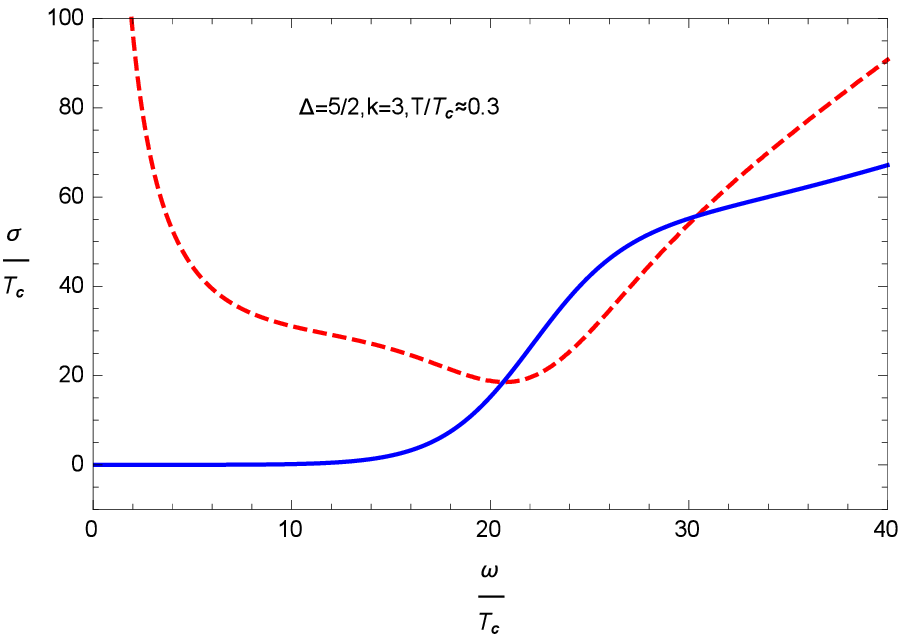}
\end{minipage}%
\begin{minipage}[c]{0.3\textwidth}
\centering\includegraphics[width=1\textwidth]{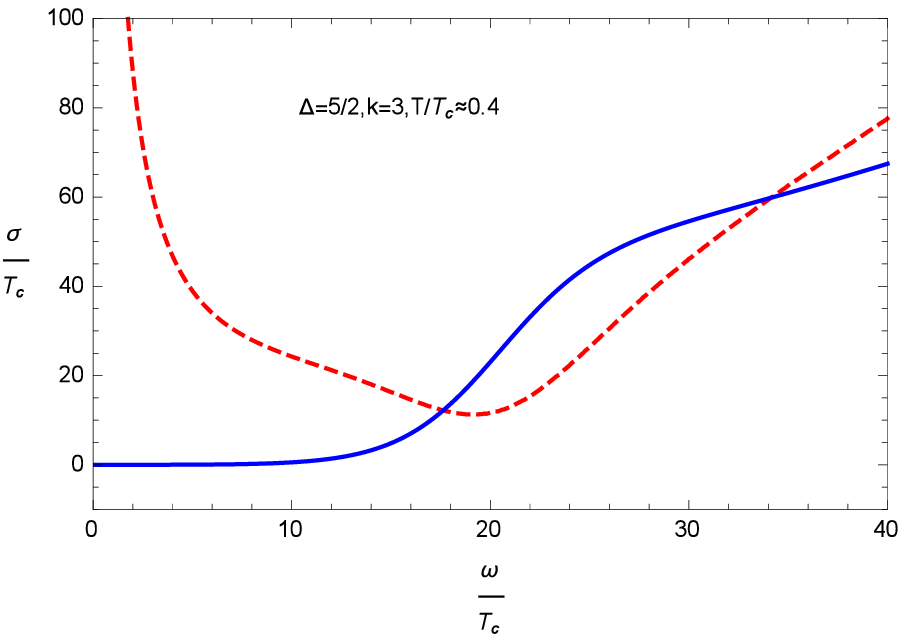}
\end{minipage}%
\begin{minipage}[c]{0.3\textwidth}
\centering\includegraphics[width=1\textwidth]{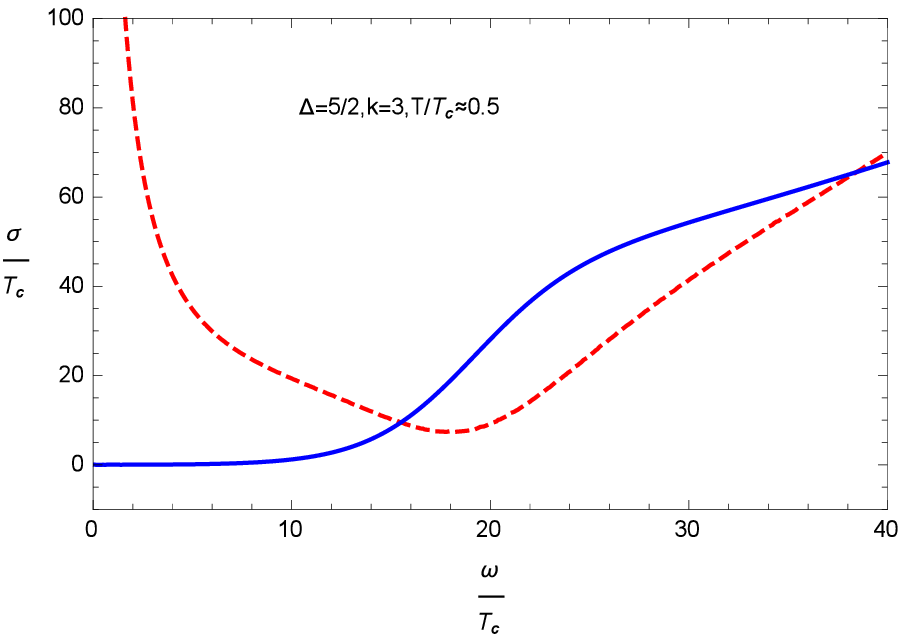}
\end{minipage}
\\
\begin{minipage}[c]{0.3\textwidth}
\centering\includegraphics[width=1\textwidth]{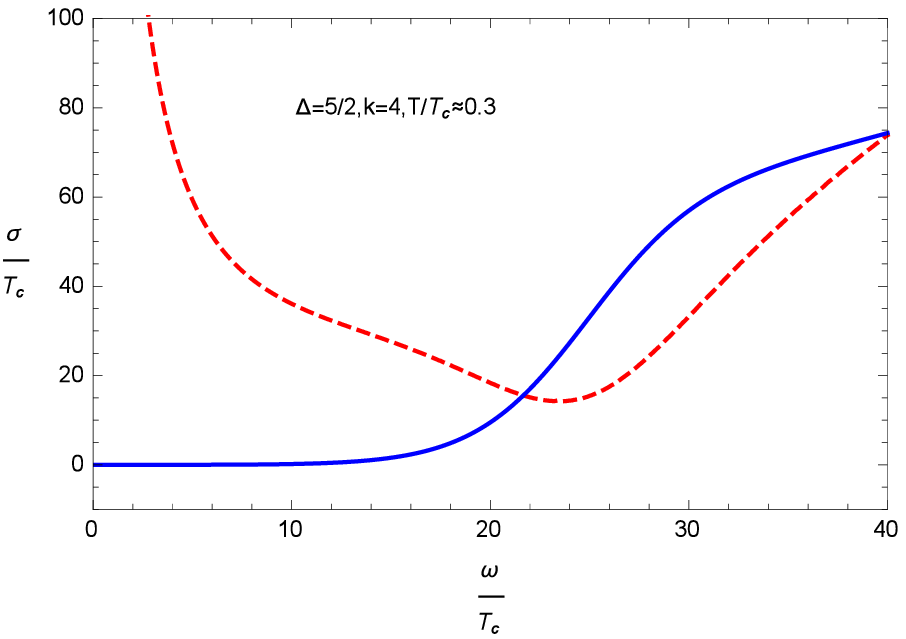}
\end{minipage}%
\begin{minipage}[c]{0.3\textwidth}
\centering\includegraphics[width=1\textwidth]{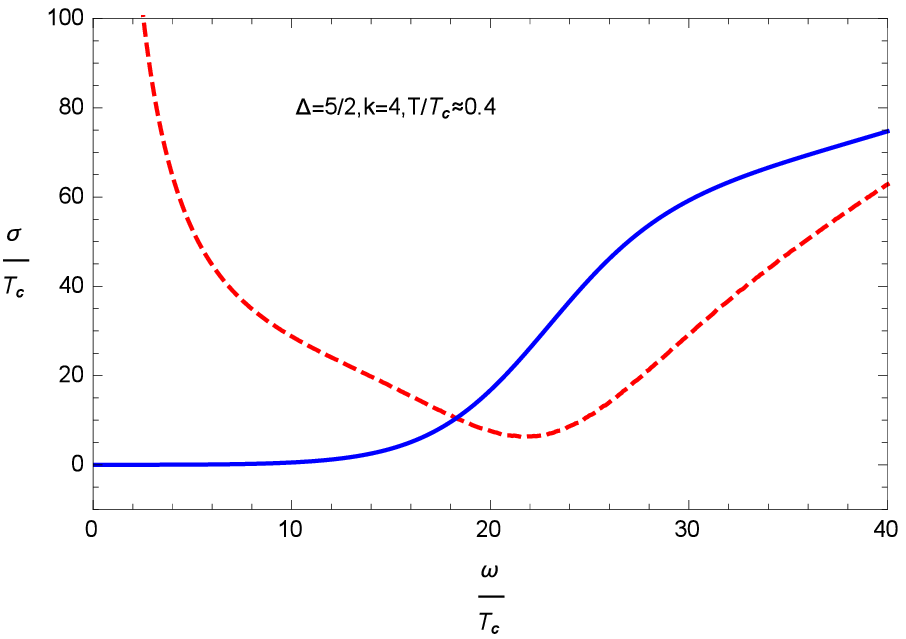}
\end{minipage}%
\begin{minipage}[c]{0.3\textwidth}
\centering\includegraphics[width=1\textwidth]{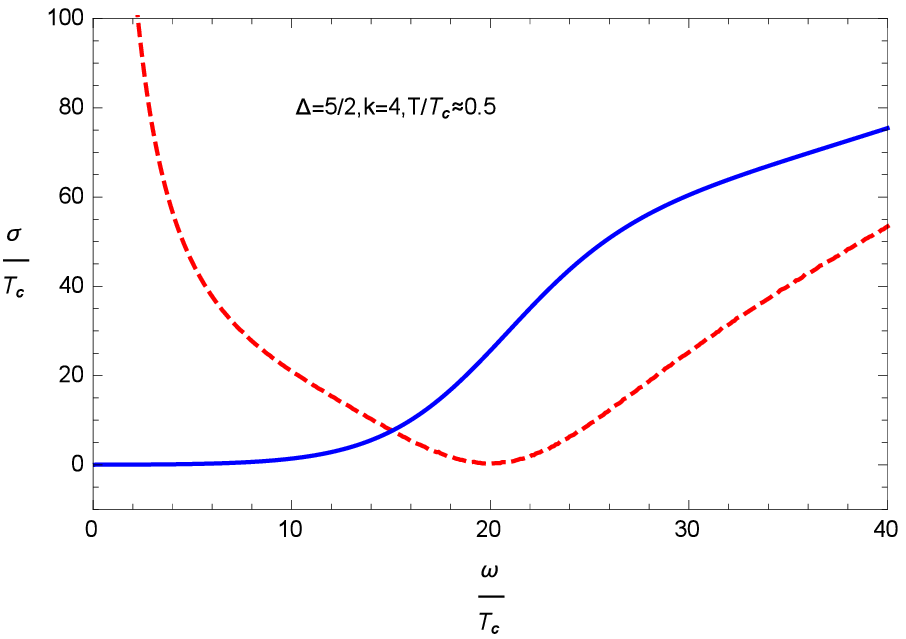}
\end{minipage}
\\
\begin{minipage}[c]{0.3\textwidth}
\centering\includegraphics[width=1\textwidth]{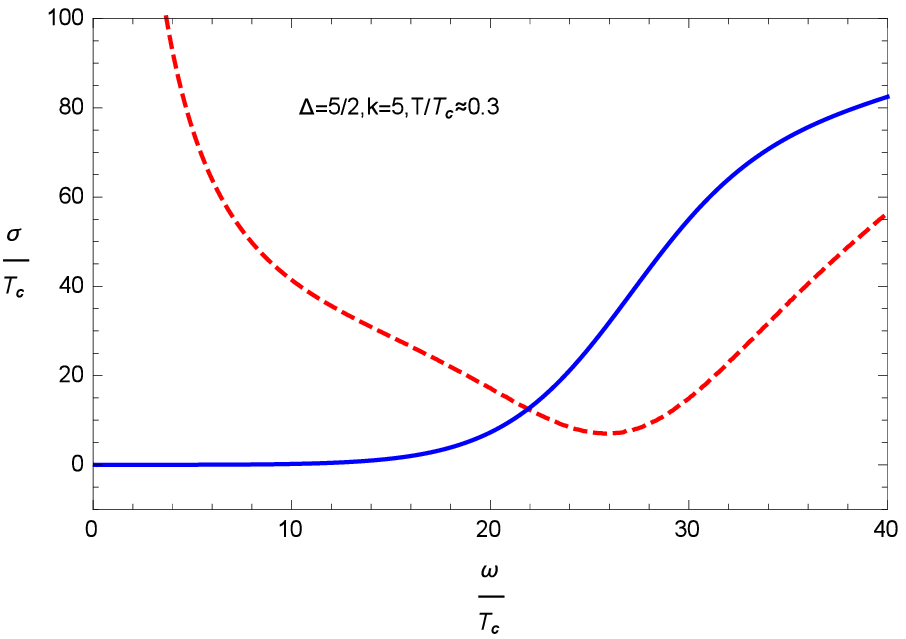}
\end{minipage}%
\begin{minipage}[c]{0.3\textwidth}
\centering\includegraphics[width=1\textwidth]{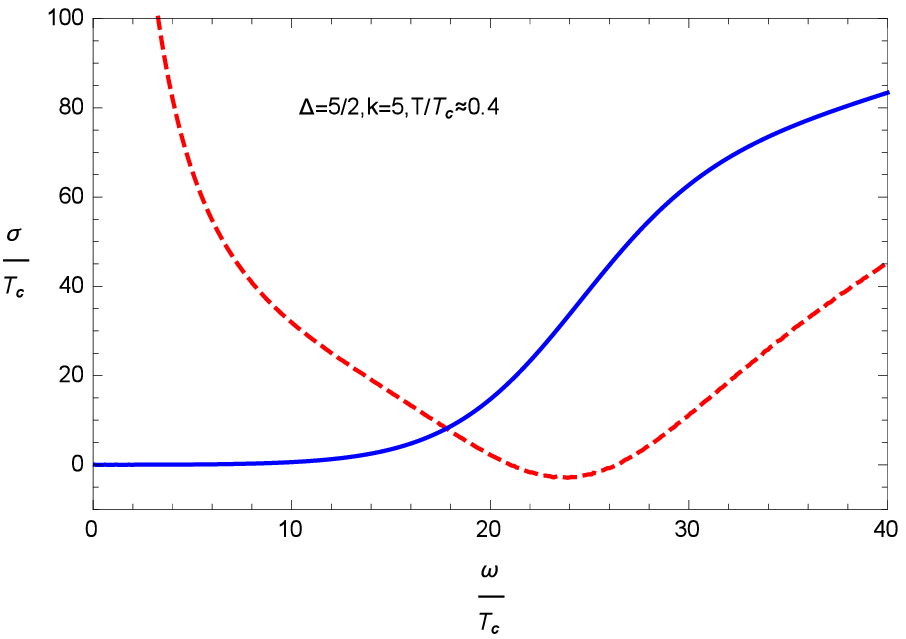}
\end{minipage}%
\begin{minipage}[c]{0.3\textwidth}
\centering\includegraphics[width=1\textwidth]{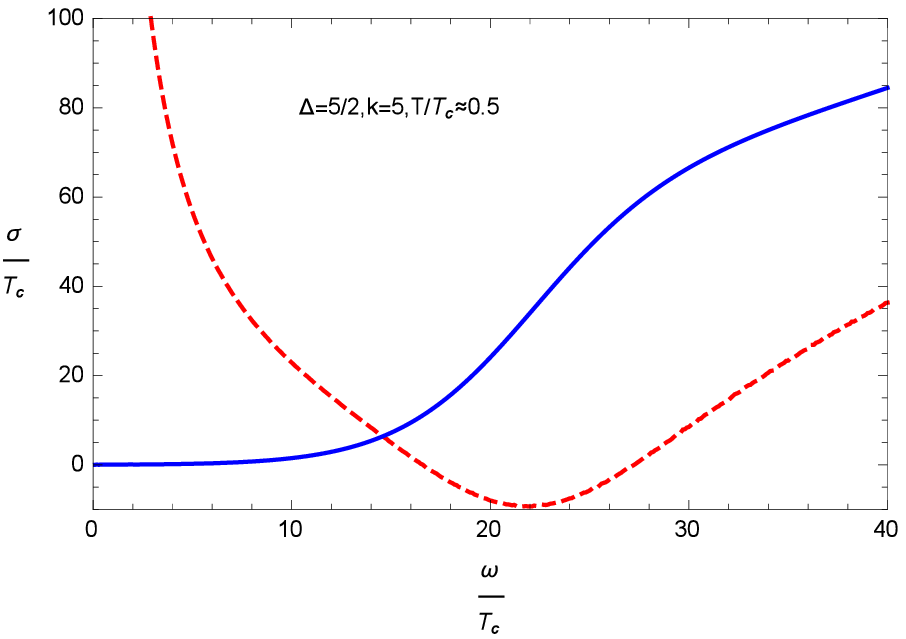}
\end{minipage}

\renewcommand{\figurename}{Fig.}
\caption{(Color online) The calculated conductivity of the holographic superconductor.
The real and the imaginary parts of the conductivity are shown in solid blue and dashed red curves.
The calculations are carried out for $\Delta=5/2, m^2=5/4$.}
\label{conductivity3}
\end{center}
\end{figure}

\begin{figure}[ht]
\begin{center}

\begin{minipage}[c]{0.3\textwidth}
\centering\includegraphics[width=1\textwidth]{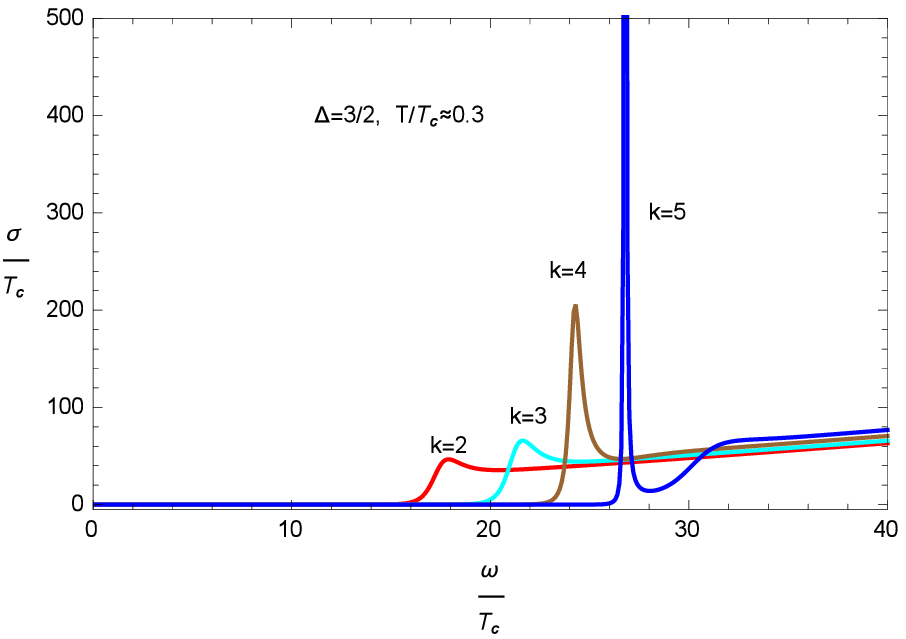}
\end{minipage}%
\begin{minipage}[c]{0.3\textwidth}
\centering\includegraphics[width=1\textwidth]{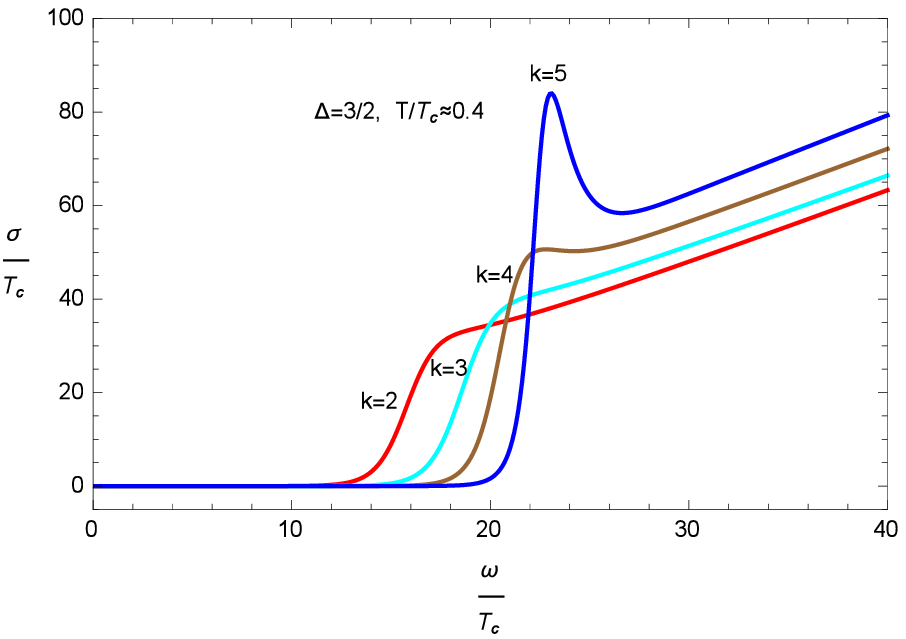}
\end{minipage}%
\begin{minipage}[c]{0.3\textwidth}
\centering\includegraphics[width=1\textwidth]{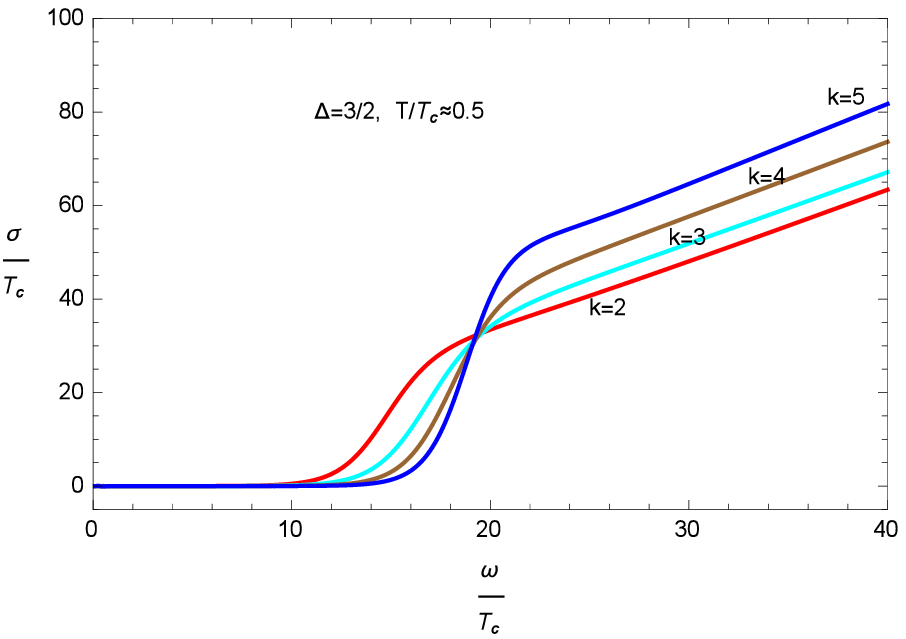}
\end{minipage}
\\
\begin{minipage}[c]{0.3\textwidth}
\centering\includegraphics[width=1\textwidth]{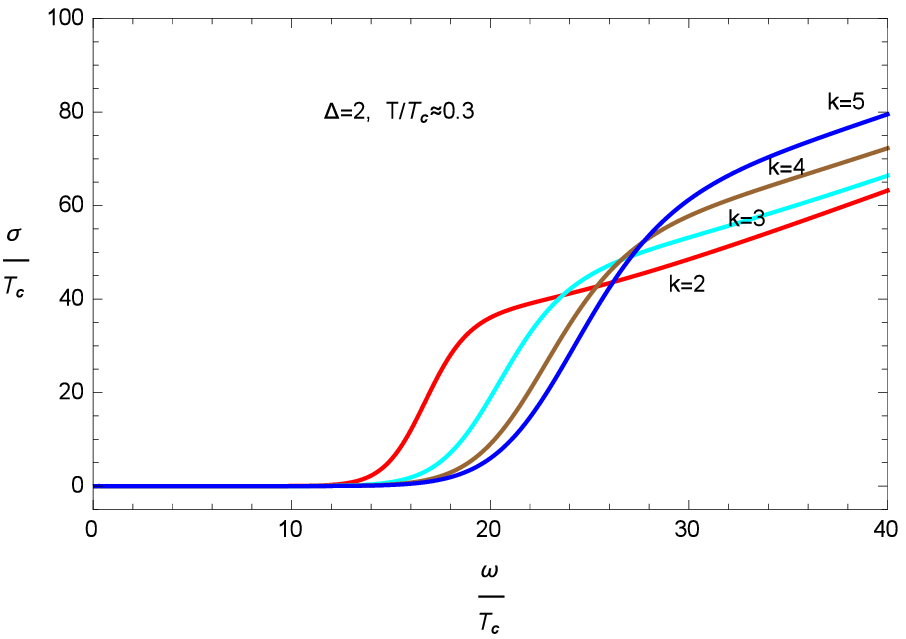}
\end{minipage}%
\begin{minipage}[c]{0.3\textwidth}
\centering\includegraphics[width=1\textwidth]{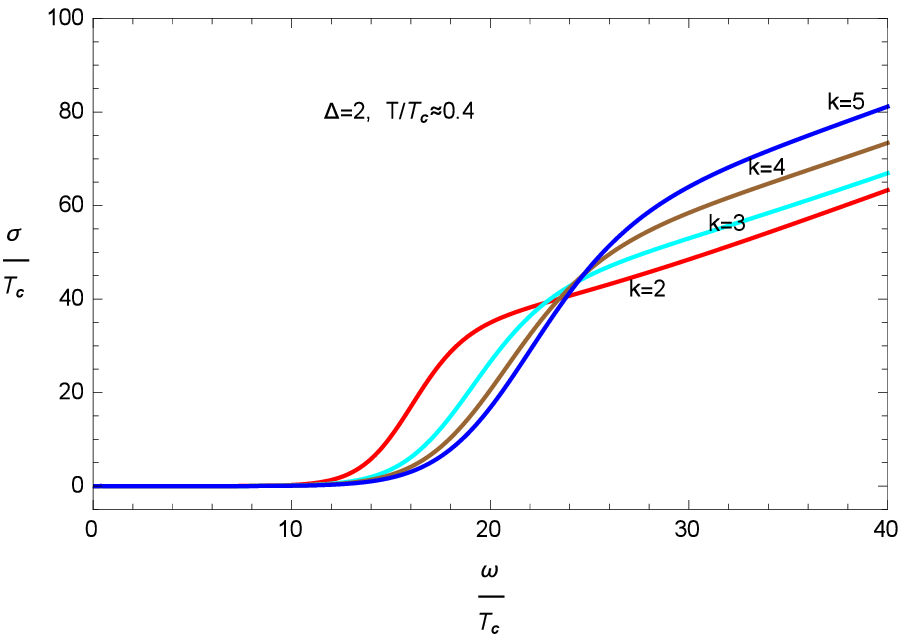}
\end{minipage}%
\begin{minipage}[c]{0.3\textwidth}
\centering\includegraphics[width=1\textwidth]{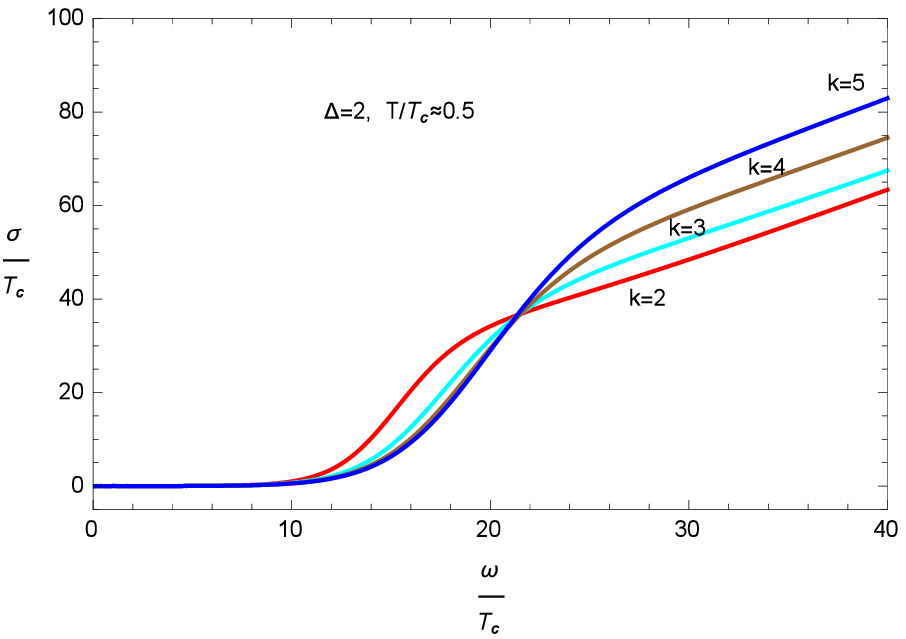}
\end{minipage}
\\
\begin{minipage}[c]{0.3\textwidth}
\centering\includegraphics[width=1\textwidth]{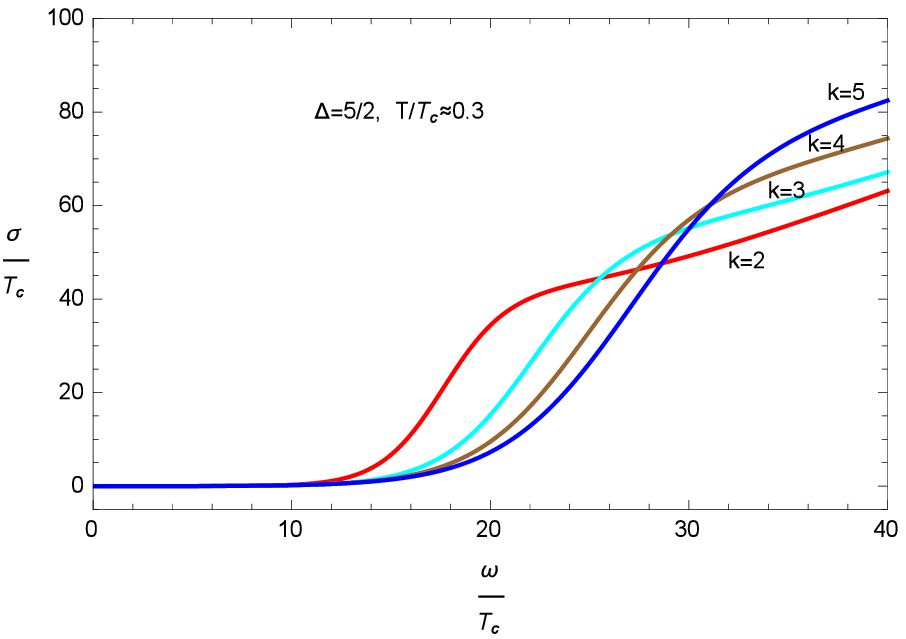}
\end{minipage}%
\begin{minipage}[c]{0.3\textwidth}
\centering\includegraphics[width=1\textwidth]{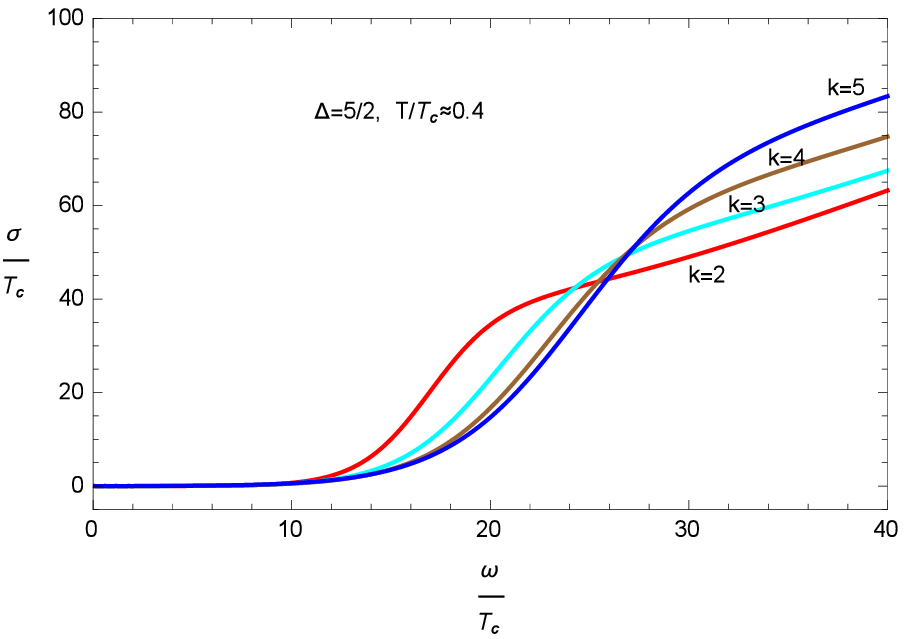}
\end{minipage}%
\begin{minipage}[c]{0.3\textwidth}
\centering\includegraphics[width=1\textwidth]{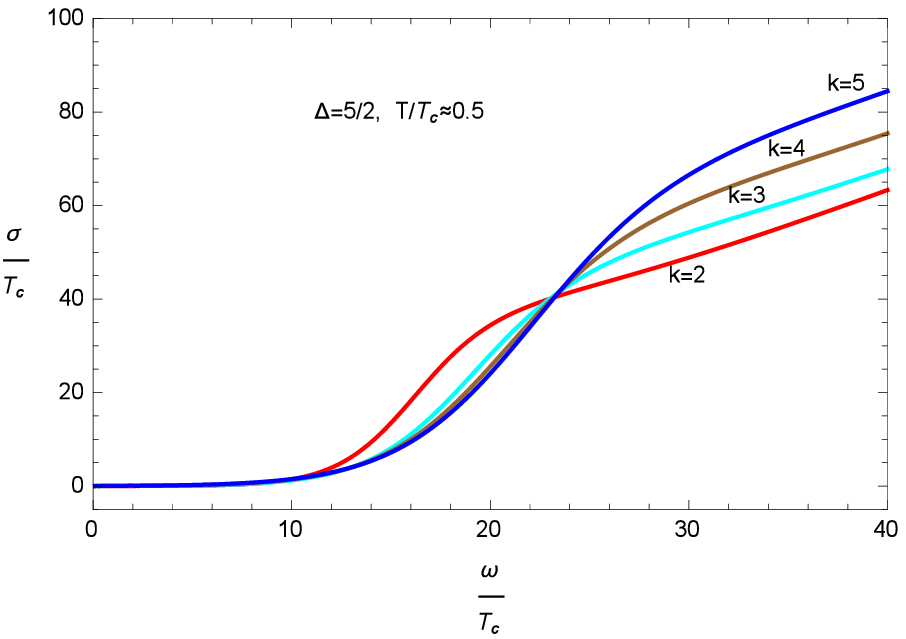}
\end{minipage}

\renewcommand{\figurename}{Fig.}
\caption{(Color online) The calculated real part of the conductivity of the holographic superconductor.
The calculations are carried out for different $\Delta, k$ and $T$.} \label{conductivity4}
\end{center}
\end{figure}

\begin{figure}[ht]
\begin{center}

\begin{minipage}[c]{0.3\textwidth}
\centering\includegraphics[width=1\textwidth]{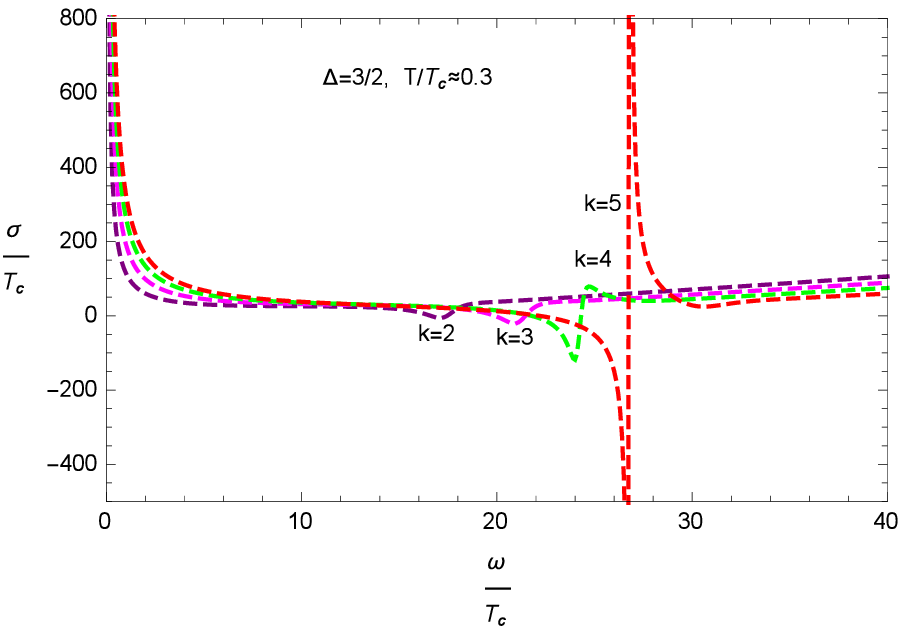}
\end{minipage}%
\begin{minipage}[c]{0.3\textwidth}
\centering\includegraphics[width=1\textwidth]{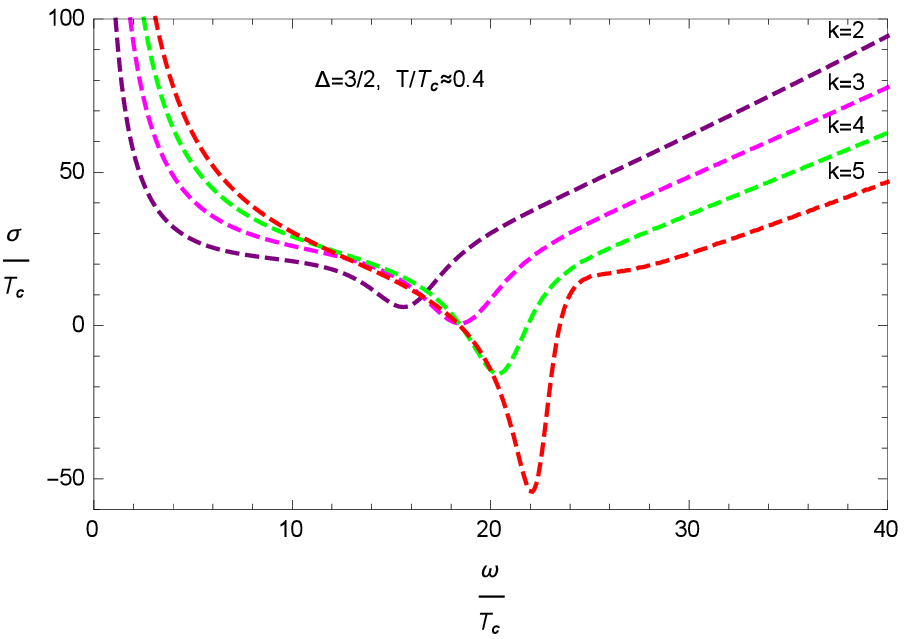}
\end{minipage}%
\begin{minipage}[c]{0.3\textwidth}
\centering\includegraphics[width=1\textwidth]{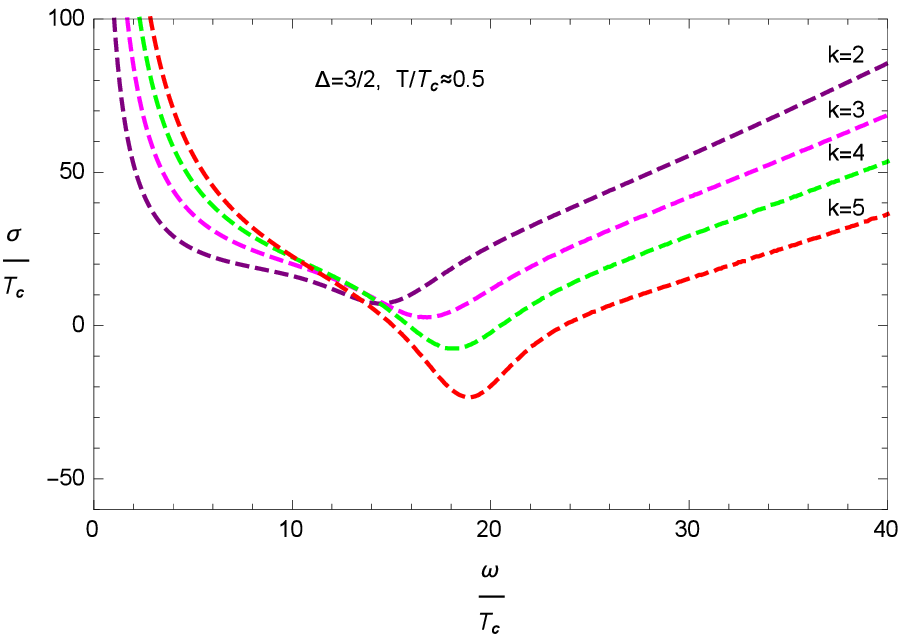}
\end{minipage}
\\
\begin{minipage}[c]{0.3\textwidth}
\centering\includegraphics[width=1\textwidth]{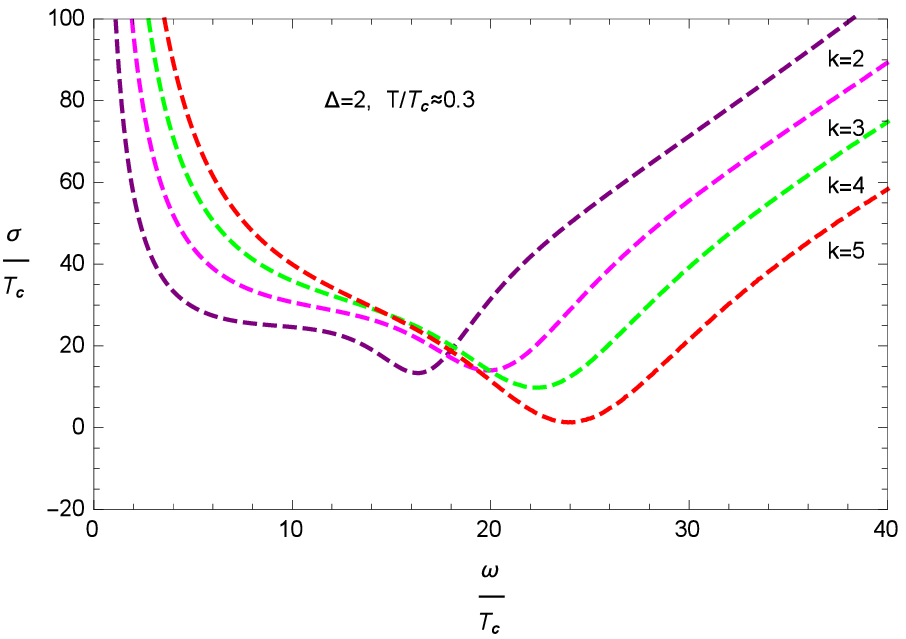}
\end{minipage}%
\begin{minipage}[c]{0.3\textwidth}
\centering\includegraphics[width=1\textwidth]{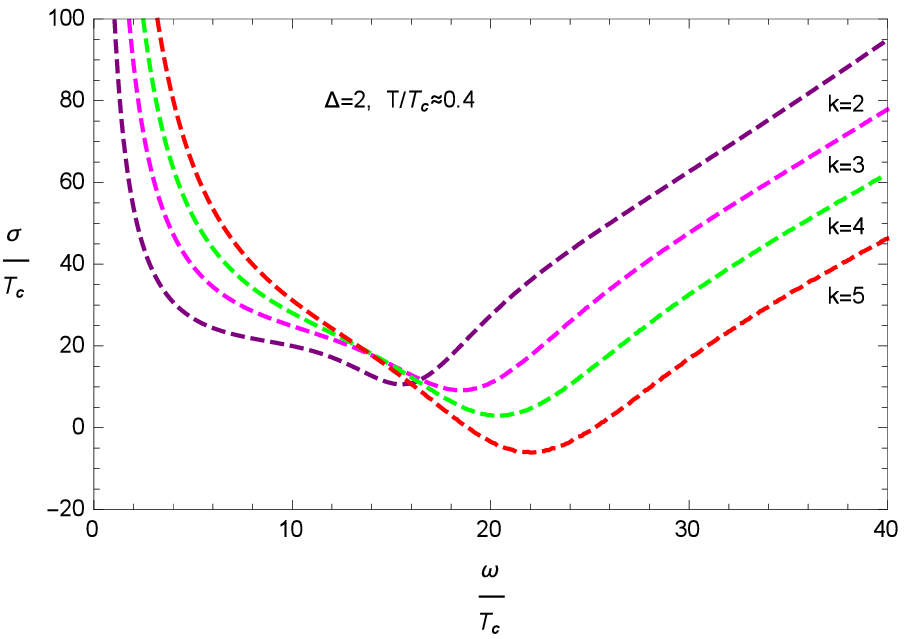}
\end{minipage}%
\begin{minipage}[c]{0.3\textwidth}
\centering\includegraphics[width=1\textwidth]{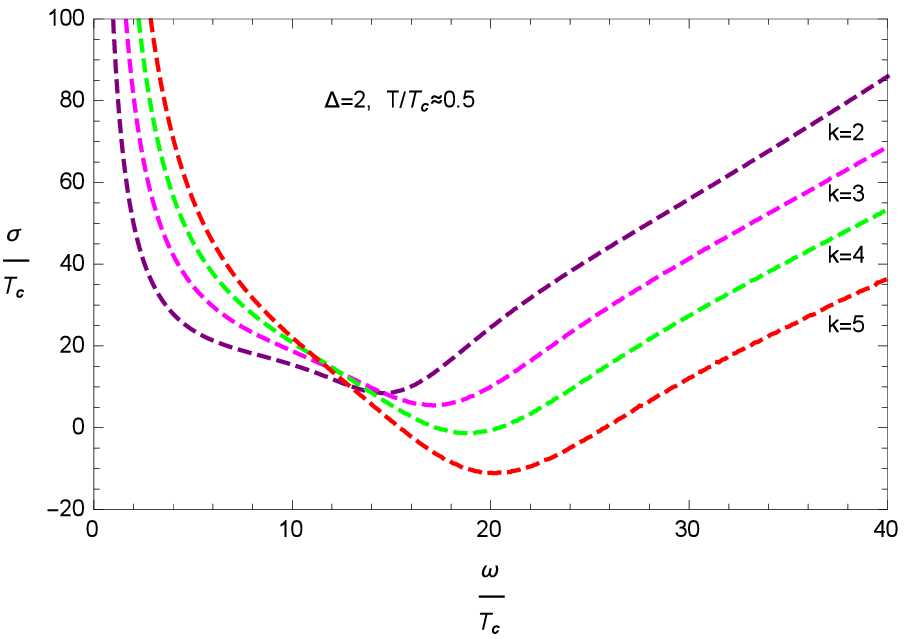}
\end{minipage}
\\
\begin{minipage}[c]{0.3\textwidth}
\centering\includegraphics[width=1\textwidth]{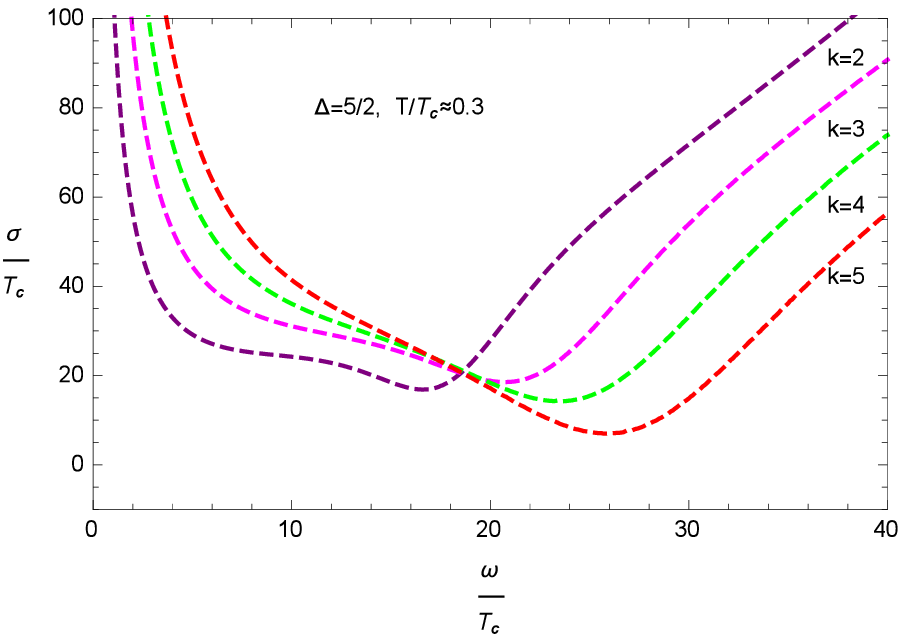}
\end{minipage}%
\begin{minipage}[c]{0.3\textwidth}
\centering\includegraphics[width=1\textwidth]{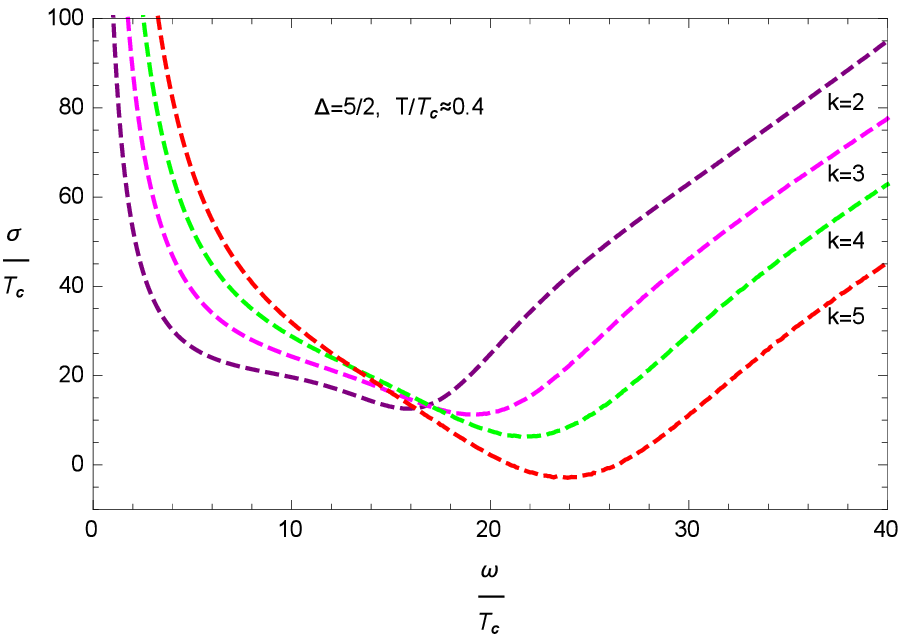}
\end{minipage}%
\begin{minipage}[c]{0.3\textwidth}
\centering\includegraphics[width=1\textwidth]{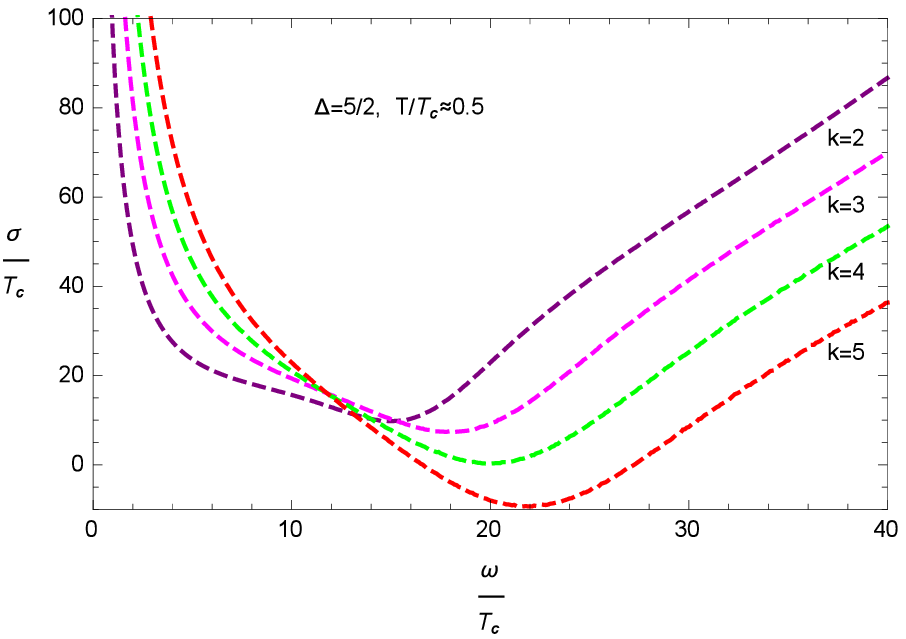}
\end{minipage}

\renewcommand{\figurename}{Fig.}
\caption{(Color online) The calculated imaginary part of the conductivity of the holographic superconductor.
The calculations are carried out for different $\Delta, k$ and $T$.}
\label{conductivity5}
\end{center}
\end{figure}

By solving Eq.(\ref{eqAy}) numerically, one obtains the conductivity as a function of the ratio between the frequency to the temperature for the various cases which are presented in Figs.\ref{conductivity1}-\ref{conductivity5}.
In Figs.\ref{conductivity1}-\ref{conductivity3}, the real part of the conductivity is compared to the imaginary part.
In the plots, the solid blue curves represent the real part, and the dashed red curves represent the imaginary part of the conductivity respectively.
To analyze its dependence on $ k$ with given $\Delta$ and $T/T_c$, we reorganize the results and show in Figs.\ref{conductivity4} and \ref{conductivity5} the real and imaginary parts of the frequency respectively.

Following \cite{GM08,LWCLRL16}, we define $\omega_g$ as the frequency that minimizes the imaginary part of the conductivity when $\Delta>\Delta_{BF}$ and $m^2>m_{BF}^2$, and evaluate the ratio of $\omega_g$ to the critical temperature $\omega_g/T_c$.
From these figures, we find that $\omega_g/T_c$ does not remain the same for different parameters, but rather changes with the variation of $\Delta$ and $k$ in this model.
Nevertheless, it turns out to be much larger than the BCS value $3.5$.

As we can see from Fig.\ref{conductivity1}, for $\Delta=3/2, m^2=-3/4$ and $T/T_c\approx 0.3$, the conductivity rises sharply at $\omega_g/T_c$, the peak of the real part increases with the increasing $k$, and the pole at $\omega_g/T_c$ gets closer to the real axis.
Especially for $k=5$ near $T/T_c\approx 0.3$, we also observe some interesting phenomena near the Breitenlohner-Freedman bound at $\Delta=\Delta_{BF}$, $m^2=m_{BF}^2$ and $T\rightarrow 0$ \cite{GM08}.
In this case, not only a second pole exists, but also it tends to move towards the real axis, which indicates that the interactions between the quasi-particles become stronger, leading to the formation of a bound state.
Furthermore, the conductivity also rises quickly near $T/T_c \approx 0.4$ and $T/T_c \approx 0.5$, whereas the phenomenon of pole hitting the real axis disappears for higher values of $T/T_c$.
On account of the above, perhaps there are more special features for a larger value of $k$ or low temperature with fixed $\Delta=3/2, m^2=-3/4$, but our numerical code can only explore down to $T/T_c\approx 0.2$ at most.

In Figs.\ref{conductivity2} and \ref{conductivity3}, we analyze the conductivity with different masses of the vector field, for which $\Delta$ is taken to be 2 and 5/2.
We find that irrespective of the various values of $\Delta$, the shapes of the curves look similar for given values of $k$ and $T/T_c$.
It seems that when the value of the vector mass $m$ is significantly different from $m_{BF}$, it doesn't have too much impact on the shape of conductivity curves.
However, the values of $k$ and $T/T_c$ affect the conductivity as in the previous cases.

From Figs.\ref{conductivity4} and \ref{conductivity5}, one finds that the gap $\omega_g/T_c$ no longer keeps unchanged as observed in Ref.\cite{GM08}.
In fact, the gap $\omega_g/T_c$ increases distinctly with increasing $k$ for given values of $\Delta$ and $T/T_c$, which implies that a smaller value of $k$ is favorable to the superconductor phase.
Moreover, the temperature also has an impact on the gap, the value of $\omega_g/T_c$ becomes larger with increasing $T/T_c$ for given values of $\Delta$ and $k$.
In contrast, the gap $\omega_g/T_c$ is not sensitive to the value of $\Delta$, which indicates that the effect of mass $m$ is comparatively smaller than those of $k$ and $T/T_c$ when $m$ is much larger than $m_{BF}$.
We also find that the peaks of the real part of conductivity increase when $k$ increases for $\Delta=3/2, T/T_c\approx 0.3$, there are poles at $\omega=0$, and both the real and imaginal parts of the conductivity diverge when the frequency $\omega\rightarrow\infty$.
Last but not least, for all these cases we study, the second pole emerges only when $\Delta=3/2, T/T_c=0.3, k=4,5$.

\section{Discussions and Conclusions}

In this paper, the properties of the p-wave superconductor in the background of higher dimensional planar AdS black holes are investigated.
What merits our primary concern is that the black hole metric is derived from a self-interacting scalar field nonminimally coupled to a particular Lovelock gravity \cite{MM13}.
This kind of black hole solutions with planar event horizon topology is characterized by an integer index number $k$.
The physical significance of the nonminimal coupling is that it might cause the black hole to form hair in addition to the mass, charge and angular momentum.
As a matter of fact, studies on the thermodynamics of a black hole solution with an arbitrary nonminimal coupling show that an integration constant appearing in the black hole solution can be interpreted as a hair since it is not associated with any conserved charge.
This work involves an attempt to investigate the effect of the higher dimensional black hole metric on the holographic superconductor.
Therefore, the focus is given to the impact of the integer index $k$ on the conductor/superconductor phase transition.
In addition, we also discussed the influences of the mass of the vector field, the spacetime dimension, and the temperature.

For the p-wave conductor/superconductor phase transition in the five-dimensional and six-dimensional cases, it is observed that the different black hole solutions, labeled by $k$, indeed have a non-negligible difference.
As $k$ increases, we found that the critical temperature decreases, the condensation gap becomes larger.
This implies that a larger parameter $k$ hinders the phase transition.
Moreover, when $\Delta$ (related to the mass of the vector field) increases, the critical temperature decreases, meanwhile the condensation gap increases.
This means that the larger the mass, the more difficult it is for the transition to take place.
Additionally, for $\Delta=3/2$ in the five-dimensional spacetime, the vector condensate tends to be insensitive to the temperature, and therefore the condensate is close to a constant below the critical temperature.
This constant value increases with the integer index $k$.
On the contrary, as $\Delta$ increases to 5/2, the condensation curve for a larger integer $k=5$ is found to tend to diverge as temperature approaches zero.
Moreover, in the six-dimensional case, for given vector field mass and $k$, the critical temperature is higher while the condensation gap is smaller than that in the five-dimensional one, and therefore the transition is easier for the former case.

The calculations of the conductivity were also carried out.
It is found that the value of $\omega_g/T_c$ varies as a function of $k$, $T/T_c$ and $\Delta$, which is different from previous findings in Ref.\cite{LWCLRL16}.
Nonetheless, the obtained value is still much larger than that achieved by the BCS theory, which suggests that the dual holographic superconductor involves strong interactions.
Another interesting feature is that a second pole in the vicinity of the real axis of $\omega/T_c$ is observed near the Breitenlohner-Freedman bound.
Numerically, a second pole appears for $\Delta=3/2>\Delta_{BF}$ and $k=2$, $T/T_c\approx 0.3$, it moves towards the real axis as $k$ increases, and gets very close for $k=5$, which indicates that another stable quasiparticle state might be formed.
This is consistent with Kramers-Kronig relations, as numerical calculations also found that the peak of the real part of conductivity sharply increases with increasing $k$, as a $\delta$ function is starting to form at the same location of the pole.
The peak decreases with increasing temperature.
These properties were observed before in an Abelian Higgs model in the AdS Schwarzschild black hole\cite{GM08}.
On the other hand, the difference between $\Delta=2$ and $\Delta=5/2$ is very small for given values of $k$ and $T/T_c$.
To sum up, we found that the integer $k$, which labels different gravity solutions, is closely related to the conductor/superconductor phase transition.
It is worthwhile to carry out the calculation by using a relatively large $k$ or at low temperature with a small $\Delta$ to further explore the properties of the system near the Breitenlohner-Freedman bound.

\begin{acknowledgments}

We are thankful for valuable discussions with Yunqi Liu, Xiaomei Kuang, Carlos Eduardo Pellicer de Oliveira, and Chenyong Zhang. 
We gratefully acknowledge the financial support from Brazilian funding agencies Funda\c{c}\~ao de Amparo \`a Pesquisa do Estado de S\~ao Paulo (FAPESP), Conselho Nacional de Desenvolvimento Cient\'{\i}fico e Tecnol\'ogico (CNPq), Coordena\c{c}\~ao de Aperfei\c{c}oamento de Pessoal de N\'ivel Superior (CAPES), and National Natural Science Foundation of China (NNSFC) under contract Nos.11690034, 11775076, 11573022 and 11375279, as well as Hunan Provincial Natural Science Foundation of China under Grant No. 2016JJ1012.

\end{acknowledgments}


\begin{thebibliography}{99}



\bibitem{G11}
G.T. Horowitz,
Notes Phys. \textbf{828} (2011) 313.

\bibitem{C09}
C.P. Herzog,
J. Phys. A \textbf{42} (2009) 343001.

\bibitem{NHM14}
N. Iqbal, H. Liu, M. Mezei,
arXiv:1110.3814 [hep-th]

\bibitem{D04}
D. Musso,
arXiv: 1401.1504 [hep-th]

\bibitem{J98}
J.M. Maldacena,
Adv. Theor. Math. Phys. \textbf{2} (1998) 231.

\bibitem{SIA98}
S.S. Gubser, I.R. Klebanov, A.M. Polyakov,
Phys. Lett. B \textbf{428} (1998) 105.

\bibitem{SCG0895}
S.A. Hartnoll, C.P. Herzog, G.T. Horowitz,
Phys. Rev. Lett. \textbf{101} (2008) 031601.

\bibitem{SCG0863}
S.A. Hartnoll, C.P. Herzog, G.T. Horowitz,
J High Energy Phys. \textbf{0812} (2008) 015.

\bibitem{GM08}
G.T. Horowitz, M.M. Roberts,
Phys. Rev. D \textbf{78} (2008) 126008.

\bibitem{GM09}
G.T. Horowitz, M.M. Roberts,
J. High Energy Phys. \textbf{0911} (2009) 015.

\bibitem{RLLR15}
R.G. Cai, L. Li, L.F. Li, R.Q. Yang,
Sci. China Phys. Mech. Astron. \textbf{58} (2015) 060401.

\bibitem{LWCLRL16}
J.W. Lu, Y.B. Wu, T. Cai, H.M. Liu, Y.S. Ren, and M.L. Liu,
Nucl. Phys. B \textbf{903} (2016) 360.

\bibitem{SS08}
S.S. Gubser, S.S. Pufu,
J. High Energy Phys. \textbf{0811} (2008) 033,

\bibitem{RSLL13}
R.G. Cai, S. He, L. Li, L.F. Li,
J. High Energy Phys. \textbf{1312} (2013) 036,

\bibitem{AY00}
A.P. Mackenzie, Y. Maeno,
Physica B \textbf{280} (2000) 148

\bibitem{RLL14}
R.G. Cai, L. Li, L.F. Li,
J. High Energy Phys. \textbf{1401} (2014) 032.

\bibitem{FDJ11}
F. Aprile, D. Rodriguez-Gomez, J.G. Russo,
J High Energy Phys. \textbf{1101} (2011) 056.

\bibitem{AJ11}
A. Donos, J.P. Gauntlett,
Energy Phys. \textbf{1112} (2011) 091.

\bibitem{AJ12}
A. Donos, J.P. Gauntlett,
Phys. Rev. Lett. \textbf{108} (2012) 211601.

\bibitem{JYDWC10}
J.W. Chen, Y.J. Kao, D. Maity, W.Y. Wen, C.P. Yeh,
Phys. Rev. D \textbf{81} (2010) 106008.

\bibitem{FCR10}
F. Benini, C.P. Herzog, R. Rahman, et al.
J High Energy Phys. \textbf{1011} (2010) 137.

\bibitem{LQJ15}
L. Zhang, Q.Y. Pan, J.L. Jing,
Phys. Lett. B \textbf{743} (2015) 104.


\bibitem{LRLC15}
L.F. Li, R.G. Cai, L. Li, C. Shen,
Nucl. Phys. B \textbf{894} (2015) 15.

\bibitem{RLLR14}
R.G. Cai, L. Li, L.F. Li, R.Q. Yang,
J. High Energy Phys. \textbf{1404} (2014) 016.

\bibitem{EL20}
E. Kiritsis, L. Li,
arXiv:1510.00020 [cond-mat.str-el].

\bibitem{YJWCJF14}
Y.B. Wu, J.W. Lu, W.X. Zhang, C.Y. Zhang, J.B. Lu, F. Yu,
Phys. Rev. D \textbf{90(12)} (2014) 126006.

\bibitem{YJCNXZS14}
Y.B. Wu, J.W. Lu, C.Y. Zhang, N. Zhang, X. Zhang, Z.Q. Yang, S.Y.
Wu,
Phys. Lett. B \textbf{741} (2014) 138.

\bibitem{CQJY16}
C.Y. Lai, Q.Y. Pan, J.L. Jing, Y.J. Wang,
Phys. Lett. B \textbf{757} (2016) 65.

\bibitem{SQJ17}
S.C. Liu, Q.Y. Pan, J.L. Jing,
Phys. Lett. B \textbf{765} (2017) 91.

\bibitem{SWY17}
S.Q. Lan, W.B. Liu, Y. Tian,
Phys. Rev. D. \textbf{95} (2017) 066013.

\bibitem{CPD09}
C.P. Herzog, P.K. Kovtun, D.T. Son,
Phys. Rev. D. \textbf{79} (2009) 066002.









\bibitem{KAQE16}
K. Lin, A.B. Pavan , Q.Y. Pan, E. Abdalla,
Braz. J. Phys. \textbf{46} (2016) 767.

\bibitem{KAQE15}
K. Lin, A.B. Pavan , Q.Y. Pan, E. Abdalla,
arXiv:1512.02718 [gr-qc]

\bibitem{KEA15}
K. Lin, E. Abdalla, A.Z. Wang,
Int. J. Mod. Phys. D \textbf{24} (2015) 1550038.

\bibitem{KE14}
K. Lin, E. Abdalla,
Eur. Phys. J. C \textbf{74} (2014) 3144.

\bibitem{MTWA09}
M. Guica, T. Hartman, W. Song, A. Strominger,
Phys. Rev. D. \textbf{80} (2009) 124008.

\bibitem{S14}
S.A. Hartnoll,
Class. Quant. Grav. \textbf{26} (2009) 224002.


\bibitem{YQY17}
Y. Peng, Q,Y. Pan, Y.Q. Liu,
Nucl. Phys. B \textbf{915} (2017) 69-83.

\bibitem{QS16}
Q.Y. Pan, S.J. Zhang,
Eur. Phys. J. C \textbf{76} (2016) 126.

\bibitem{QJBS14}
Q.Y. Pan, J.L. Jing, B. Wang, S.B. Chen,
J. High Energy Phys. \textbf{1206} (2012) 087.

\bibitem{romansan}
R. A. Konoplya and A. Zhidenko,
Phys. Lett. \textbf{B686}, (2010), 199.




\bibitem{QJB11}
Q.Y. Pan, J.L. Jing, B. Wang,
J. High Energy Phys. \textbf{1111} (2011) 088.

\bibitem{SZ13}
S.L. Cui, Z. Xue,
Phys. Rev. D \textbf{88(10)} (2013) 107501.

\bibitem{WJ13}
W.P. Yao, J.L. Jing,
J. High Energy Phys. \textbf{1305} (2013) 101.

\bibitem{XLR14}
X.Y. Guo, L.C. Zhang, R. Zhao,
Mod. Phys. Lett. A \textbf{29(19)} (2014) 1450083.

\bibitem{OL}
O. Miskovic, L. Aranguiz,
The Thirteenth Marcel Grossmann Meeting: pp.1425.

\bibitem{ZH15}
Z.Y. Nie, H. Zeng,
J. High Energy Phys. \textbf{1510} (2015) 047,

\bibitem{DS44}
D. Ghorai, S. Gangopadhyay,
arXiv:1511.02444 [hep-th].

\bibitem{GY15}
G. Liu, Y. Peng,
Mod. Phys. Lett. A \textbf{30} (2015) 1550183.

\bibitem{R02}
R.G. Cai,
Phys. Rev. D \textbf{65} (2002) 084014.

\bibitem{RSB07}
R.G. Cai, S.P. Kim, B. Wang,
Phys. Rev. D \textbf{76} (2007) 024011.

\bibitem{RSJ09}
R. Gregory, S. Kanno, J. Soda,
J. High Energy Phys. \textbf{0910} (2009) 010.

\bibitem{RZH10}
R.G. Cai, Z.Y. Nie, H.Q. Zhang,
Phys. Rev. D \textbf{82} (2010) 066007.

\bibitem{QB10}
Q.Y. Pan, B. Wang,
Phys. Lett. B \textbf{693} (2010) 159.

\bibitem{HRH11}
H.F. Li, R.G. Cai, H.Q. Zhang,
J. High Energy Phys. \textbf{1104} (2011) 028.

\bibitem{RZH11}
R.G. Cai, Z.Y. Nie, H.Q. Zhang,
Phys. Rev. D \textbf{83} (2011) 066013.

\bibitem{QBEJA10}
Q.Y. Pan, B. Wang, E. Papantonopoulos, J. Oliveira, A.B. Pavan,
Phys. Rev. D \textbf{81} (2010) 106007.

\bibitem{SD12}
S. Gangopadhyay, D. Roychowdhury,
J. High Energy Phys. \textbf{1205} (2012) 156.

\bibitem{TST10}
T.Nishioka, S. Ryu, T.Takayanagi,
J High Energy Phys. \textbf{1003} (2010) 131.

\bibitem{PG15}
P. Chaturvedi, G. Sengupta,
J. High Energy Phys. \textbf{1504} (2015) 001.

\bibitem{CG08}
C. Garraffo, G. Giribet,
Mod. Phys. Lett. A \textbf{23} (2008) 1801.

\bibitem{RA13}
R. Brustein, A.J.M. Medved,
Phys. Rev. D \textbf{88} (2013) 064010

\bibitem{MM13}
M.B. Gaete, M. Hassa\"{i}ne,
J. High Energy Phys. \textbf{11} (2013) 177.

\bibitem{FM14}
F. Correa, M. Hassa\"{i}ne,
J. High Energy Phys. \textbf{02} (2014) 014

\bibitem{UC-01}
K. Maeda, M. Natsuume, and T. Okamura, Phys. Rev. D \textbf{79}
(2009) 126004.

\bibitem{MM13prd}
M.B. Gaete, M. Hassa\"{i}ne,
Phys.Rev. D. \textbf{88} (2013) 104011.

\bibitem{JLJ57}
J. Bardeen, L.N. Cooper, J.R. Schrieffer.
Phys Rev. \textbf{108} (1957) 1175


\end{thebibliography}
\end{document}